\documentclass[traditabstract]{aa}
\usepackage{natbib,graphicx}
\usepackage{amsmath,amsfonts,amssymb}
\usepackage{txfonts}
\usepackage{longtable}
\usepackage{soul,color,lscape}
\usepackage[breaklinks,colorlinks,citecolor=blue]{hyperref}

\newcommand{\aaa}[1]{#1}
\newcommand{\corr}[1]{#1}

\begin{document}

\title{\textit{Euclid} preparation: IV. Impact of undetected galaxies on weak-lensing shear measurements \thanks{Based on {\it Hubble} Space Telescope Ultra-Deep Field (HST-UDF) data.}}

\titlerunning{Impact of undetected galaxies on weak-lensing shear measurements}
\authorrunning{Euclid Collaboration}

\author{\textit{Euclid} Collaboration, N.~Martinet$^{1,2}$, T.~Schrabback$^{1}$, H.~Hoekstra$^{3}$, M.~Tewes$^{1}$, R.~Herbonnet$^{4}$, P.~Schneider$^{1}$, B.~Hernandez-Martin$^{1}$, A.N.~Taylor$^{5}$, J.~Brinchmann$^{3,6}$, C.S.~Carvalho$^{7}$, M.~Castellano$^{8}$, G.~Congedo$^{5}$, B.R.~Gillis$^{5}$, \smash{E.~Jullo$^{2}$, M.~K\"{u}mmel$^{9}$, S.~Ligori$^{10}$, P.B.~Lilje$^{11}$, C.~Padilla$^{12}$, D.~Paris$^{8}$, J.A.~Peacock$^{5}$, S.~Pilo$^{8}$, A.~Pujol$^{13,14}$,} D.~Scott$^{15}$, R.~Toledo-Moreo$^{16}$}

\institute{$^{1}$ Argelander-Institut f\"ur Astronomie, Universit\"at Bonn, Auf dem H\"ugel 71, 53121 Bonn, Germany\\
$^{2}$ Aix-Marseille Univ, CNRS, CNES, LAM, Marseille, France\\
$^{3}$ Leiden Observatory, Leiden University, Niels Bohrweg 2, 2333 CA Leiden, the Netherlands\\
$^{4}$ Department of Physics and Astronomy, Stony Brook University, Stony Brook, NY 11794, USA\\
$^{5}$ Institute for Astronomy, University of Edinburgh, Royal Observatory, Blackford Hill, Edinburgh EH9 3HJ, UK\\
$^{6}$ Instituto de Astrof\'isica e Ci\^encias do Espa\c{c}o, Universidade do Porto, CAUP, Rua das Estrelas, PT4150-762 Porto, Portugal\\
$^{7}$ Instituto de Astrof\'isica e Ci\^encias do Espa\c{c}o, Faculdade de Ci\^encias, Universidade de Lisboa, Tapada da Ajuda, PT-1349-018 Lisboa, Portugal\\
$^{8}$ INAF-Osservatorio Astronomico di Roma, via Frascati 33, I-00078 Monteporzio Catone, Italy\\
$^{9}$ Universit\"ats-Sternwarte M\"unchen, Fakult\"at f\"ur Physik, Ludwig- Maximilians-Universit\"at M\"unchen, Scheinerstrasse 1, 81679 M\"unchen, Germany\\
$^{10}$ INAF-Osservatorio Astrofisico di Torino, via Osservatorio 20, 10025 Pino Torinese (TO), Italy\\
$^{11}$ Institute of Theoretical Astrophysics, University of Oslo, P.O. Box 1029 Blindern, N-0315 Oslo, Norway\\
$^{12}$ Institut de F\'isica d’Altes Energies IFAE, 08193 Bellaterra, Barcelona, Spain\\
$^{13}$ Universit\'e Paris Diderot, AIM, Sorbonne Paris Cit\'e, CEA, CNRS F-91191 Gif-sur-Yvette Cedex, France\\
$^{14}$ IRFU, CEA, Universit\'e Paris-Saclay F-91191 Gif-sur-Yvette Cedex, France\\
$^{15}$ Departement of Physics and Astronomy, University of British Columbia, Vancouver, BC V6T 1Z1, Canada\\
$^{16}$ Depto. de Electr\'onica y Tecnolog\'ia de Computadoras Universidad Polit\'ecnica de Cartagena, 30202, Cartagena, Spain\\
  \email{nicolas.martinet@lam.fr}}

\setcounter{page}{1}

\abstract{In modern weak-lensing surveys, \aaa{the common approach to correct for residual systematic biases in the shear is to calibrate shape measurement algorithms using simulations.} These simulations must fully capture the complexity of the observations to avoid introducing any additional bias. In this paper we study the importance of faint galaxies below the observational detection limit of a survey. We simulate simplified {\it Euclid} VIS images including \corr{and} excluding this faint population, and measure the shift in the multiplicative shear bias between the two sets of simulations. We measure the shear with three different algorithms: a moment-based approach, model fitting, and machine learning. We find that for all methods, a spatially uniform random distribution of faint galaxies introduces a shear multiplicative bias of the order of a few times $10^{-3}$. This value increases to the order of $10^{-2}$ when including the clustering of the faint galaxies, as measured in the {\it Hubble} Space Telescope Ultra-Deep Field. The magnification of the faint background galaxies due to the brighter galaxies along the line of sight is found to have a negligible impact on the multiplicative bias. We conclude that the undetected galaxies must be included in the calibration simulations with proper clustering properties down to magnitude 28 in order to reach a residual uncertainty on the multiplicative shear bias calibration of a few times $10^{-4}$, in line with the $2\times10^{-3}$ total accuracy budget required by the scientific objectives of the {\it Euclid} survey. We propose two complementary methods for including faint galaxy clustering in the calibration simulations.}


\maketitle

\section{Introduction}
\label{sec:intro}

Cosmic shear, the coherent weak lensing (WL) distortion (`shear') of galaxy images by the large-scale structure of the Universe, is one of the most powerful cosmological probes. Two particularly powerful aspects of the method are that it is based on a geometrical observable, that is, the distorted shapes of galaxy images, and that it is sensitive to the gravitational potential of structures, and as such probes both baryonic and dark matter. The usual estimator of cosmic shear is the ellipticity two-point correlation function, which quantifies the coherent distortion between pairs of galaxies as a function of their separation. Applying this estimator to recent weak-lensing surveys has yielded some of the tightest low-redshift cosmological constraints on the matter density and the amplitude of the matter power spectrum (see e.g., \aaa{\citet{Kilbinger+13, Jee+16, Hildebrandt+17, Troxel+18,Hikage+19,Chang+19})}. Complementary estimators are also in development and might require specific treatment of shear-measurement systematic errors: for example, galaxy--galaxy lensing as a two-point statistic \aaa{\citep[e.g.,][]{DES+18,Joudaki+18,vanUitert+18,Singh+18}} and the peaks in weak-lensing reconstructed mass maps as a higher-order statistic \aaa{\citep[e.g.,][]{Kacprzak+16,Martinet+18}}. \aaa{Recent reviews} of cosmic shear can be found in, \aaa{for example, \citet{Kilbinger+15,Mandelbaum18}}.

The great potential of cosmic shear has led to the development of large dedicated surveys that will gather data in the near future: {\it Euclid},\footnote{\url{https://www.euclid-ec.org/}} {\small WFIRST},\footnote{\url{https://wfirst.gsfc.nasa.gov/}} and {\small LSST}.\footnote{\url{https://www.lsst.org/}} In particular, the {\it Euclid} satellite will survey $15\,000$ deg$^2$ of the sky in order to shed light on the nature of dark energy (DE), responsible for the accelerated expansion of our Universe. This will be achieved by measuring the possible deviation of the DE equation of state parameter $w$ from the value $-1$, which corresponds to the case of a cosmological constant $\Lambda$. To reach the full statistical potential of the survey, it is mandatory to keep systematic biases on the shear measurement low. \citet{Massey+13} showed that the total multiplicative shear bias, which quantifies systematic errors in the amplitude of the shear, must be lower than $2\times10^{-3}$, and \citet{Cropper+13} presented a breakdown of this requirement over the known sources of bias, taking into account the {\it Euclid} survey and instrument design.

The amplitude of the shear due to the large-scale structure is typically a few times $10^{-2}$, which is an order of magnitude smaller than the dispersion of intrinsic galaxy ellipticities ($\sim 0.3$). These introduce shape noise, which can be mitigated by averaging the ellipticity measurements over a statistical sample of source galaxies affected by a similar distortion, assuming that galaxies have random orientations. In that case the average ellipticity yields an unbiased estimate of the mean shear. \aaa{This estimator is however biased by intrinsic alignments of galaxies in observations \citep[e.g.,][]{Hirata+04}.} The image point-spread function (PSF) also affects observed galaxy images, introducing not only blurring, but also spurious distortions that can easily exceed the cosmological shear. The PSF is corrected for by using measurements of stars, which are point-like sources in the images, or by carefully modeling it from the telescope specifications. The latter option is only possible in space-based observations, where the atmosphere does not add further deformation to the PSF.

Many methods have been proposed to carry out  measurements of galaxy shape. They can be classified into two main categories: moment measurements and model fitting. A first approach to measure the moments of the surface-brightness distribution of stars and galaxies to infer PSF-corrected estimates of galaxy ellipticities was developed by \citet{KSB95}, which is often referred to as KSB. \texttt{DEIMOS} \citep{Melchior+11} is another example of such a method, however this latter does not require the assumption of a shape for the PSF. Model-fitting methods rely on directly fitting the galaxy surface brightness profile convolved with the PSF model. These methods can yield a highly accurate correction for the PSF, but are computationally demanding as they require the difference between the model and observed profile to be minimized for every galaxy. Various model-fitting algorithms have been developed: for example, \texttt{sFIT} \citep{Jee+13}, \texttt{{\it lens}fit} \aaa{\citep{Miller+07, Miller+13}}, \aaa{\texttt{IM3SHAPE} \citep{Zuntz+13}, \texttt{NGMIX} \citep{Sheldon+14}}, and \aaa{the lensing-dedicated implementation of \texttt{SExtractor}/\texttt{PSFEx} \citep{Bertin+96,Bertin11} described in \citet{Mandelbaum+15}}. \citet{Simon+17} also recently showed that moment-based methods are similar to model-fitting with the moments being an imperfect fit to the surface-brightness distribution. Supervised machine learning, trained on image simulations, can then be used to correct for the imperfections of these measurements. \texttt{MomentsML} \citep{Tewes+19} is an example using neural networks to obtain accurate shear estimates based on moment measurements on the galaxy images. \texttt{BFD} \citep[Bayesian Fourier Domain:][]{Bernstein+14, Bernstein+16} is another moment-based refined technique, which compresses the pixel information and then estimates the probability distribution of these pixels being gravitationally distorted.

The variety of available methods has given rise to several international challenges to compare them. This started with the Shear TEsting Programmes (STEP) by blindly running the algorithms on simulated images where the input shear is compared with the output of each method \citep{Heymans+06, Massey+07}. The simulations were later modified in the GRavitational LEnsing Accuracy Testing (GREAT) challenges to check for specific effects on shear measurements, mimicking both ground-based and space-based observations \citep{Bridle+10, Kitching+12, Mandelbaum+14, Mandelbaum+15}. After these challenges, it became clear that shear measurement algorithms need to be calibrated using simulations to correct for systematic biases if one wants to reach the accuracy required by modern surveys. This is already the approach followed \aaa{by the Kilo Degree Survey \citep{FenechConti+17,Kannawadi+18}, the Dark Energy Survey \citep{Zuntz+18,Samuroff+18}, and the Hyper Suprime-Cam survey \citep{Mandelbaum+18} teams,} who created specific sets of simulations to mimic their observations and calibrate their shear measurement algorithms. We note that some of the newer methods, such as \texttt{BFD} and \texttt{METACALIBRATION} \aaa{\citep{Sheldon+17,Huff+17}}, which measures the shear response by directly distorting the observed images,  do not require calibration simulations in principle, but still \aaa{use simulations} in practice in order to \aaa{inform the prior on residual biases due to specific effects} (such as blends).

Although they allow one to correct for most systematic biases, relying on simulations means that the performance of the shape measurement algorithm will depend on how realistic these simulations are \citep[see e.g.,][]{Hoekstra+15}. Indeed, any difference between the calibration set and the observed data will introduce new biases. Many simplifications are made in the simulations and it is paramount to ensure that they do not significantly add to the original shear measurement bias breakdown of the {\it Euclid} mission \citep{Cropper+13}. Insufficiently explored simplifications include the assumption of uniform background and neglecting noise correlations \aaa{\citep[see e.g.,][]{Gurvich+16}}, which can be caused by faint undetected galaxies. The use of analytic surface brightness profiles instead of real galaxy shapes is another common simplification which began to be explored in, for example, \aaa{\citet{Lewis09,Mandelbaum+15}}. Finally, the effect of neglecting the wavelength dependence of the galaxy profile, also known as color gradient, has been studied in \aaa{\citet{Voigt+12,Semboloni+13,Er+18}, for example}.

In this paper we focus on the impact of the galaxies below the $10\sigma$ detection limit of the visible  instrument (VIS) of the {\it Euclid} mission. The VIS is an optical camera composed of 36 \corr{charge-coupled devices (CCDs)} with a field of view of $0.57~{\rm deg^2}$  covering a wavelength range from 550 to 900~nm. This detection limit corresponds to a VIS AB magnitude of 24.5, as described in \citet{Cropper+16}. These faint galaxies act as a source of correlated noise in the vicinity of the detectable galaxies, affecting both galaxy shapes and background determination, and might therefore bias their shear measurement. This question has been tackled in \citet{Hoekstra+17}, who showed using their moment-based shear-measurement algorithm that the faint galaxies need to be included down to a magnitude of about 27 to 29 in the calibration simulations in order to account for a multiplicative shear bias of the order of a few $10^{-3}$ caused by the faint galaxies and measured with an uncertainty of $\sim10^{-4}$. \aaa{\citet{Samuroff+18} also investigated the impact of faint neighboring galaxies on shear measurements for the Dark Energy Survey calibration simulations. Their approach is similar to \citet{Hoekstra+17} but for a model-fitting shape measurement algorithm (\texttt{IM3SHAPE}). Importantly, the clustering between the undetected and the detected galaxies is neglected in both studies.}

Similarly to \citet{Hoekstra+17} \aaa{and \citet{Samuroff+18}}, in this paper we investigate the bias due to the faint galaxies by comparing shear measurements in simulations with and without these galaxies. However, we improve on various aspects of the simulation of the faint population. First, we make use of the {\it Hubble} Space Telescope Ultra-Deep Field \citep[HST-UDF,\footnote{\url{http://www.stsci.edu/hst/udf}}][]{Beckwith+06} images to generate a realistic population of faint galaxies as measured from the observations down to an F775W magnitude of 29. Second, we include the clustering of the faint galaxies around bright ones, as measured in the UDF. This is expected to have a strong impact on shape measurement, since it places the faint unresolved galaxies closer to the detectable ones. It is important to note that we study only the impact of the clustering of the unresolved galaxies, and therefore isolate this effect from that of nearby resolved sources which is a separate issue \aaa{\citep[see e.g.,][for studies of the latter effect]{Samuroff+18,Mandelbaum+18}}. In physical terms, the clustered faint galaxies correspond to satellite galaxies, that is, the one-halo term in the halo model approach. We also investigate the impact of the magnification of the background faint population due to the bright galaxies along the line of sight. Finally, we generalize the measurement to three different algorithms, representative of the main classes of shape-measurement algorithms. We use a refined version of the \texttt{KSB+} method presented in \citet{Schrabback+10}, \texttt {SExtractor} and \texttt{PSFEx} for a model-fitting method, and \texttt{MomentsML} \citep{Tewes+19} for a machine-learning approach.

We describe our simulation pipeline in Sect.~\ref{sec:method}, starting from measuring galaxy populations and their clustering properties in the UDF, followed by a description of how we generate mock galaxy catalogs and their corresponding synthetic images. The three different shear-measurement methods are briefly described in Sect.~\ref{sec:shape}, and Sect.~\ref{sec:bias} summarizes how we quantify shear bias and estimate the required number of simulated galaxies. We present our results in the two subsequent  sections, where in Sect.~\ref{sec:unclus} the clustering of faint galaxies around bright galaxies is neglected. In particular, we study the \corr{magnitude limit up to} which faint galaxies impact the shear bias, the importance of getting a realistic estimate on their sizes, and stress the effects of proper background subtraction. We show in Sect.~\ref{sec:clus} that the effect of this clustering on the multiplicative bias is indeed dramatic, and study several dependences, such as the clustering length and the deblending strategy. We show in Sect.~\ref{sec:magnif} that magnification effects are only of minor importance. We then discuss in Sect.~\ref{sec:discu} the strategy that future image simulations for Euclid calibrations ought to use to account for the effects of clustering of faint galaxies, before we conclude in Sect.~\ref{sec:ccl}.

\section{Building realistic simulations}
\label{sec:method}

To quantify the effect of undetected galaxies we construct simulations including and excluding them, comparing the shape measurement of the detectable galaxies from both sets of simulations. We first build a catalog of realistic galaxies in the VIS AB magnitude range [20, 29], measuring photometric properties in the HST-UDF images. We then sample from that catalog to generate a random ensemble of galaxies with realistic properties, taking into account correlations between parameters. Finally, we use the \texttt{GalSim} software \citep{Rowe+15} to generate images of these galaxies, mimicking the observing conditions of the {\it Euclid} VIS instrument.

\subsection{Measuring galaxy properties}

\begin{figure}
\centering
\includegraphics[width=0.5\textwidth,clip]{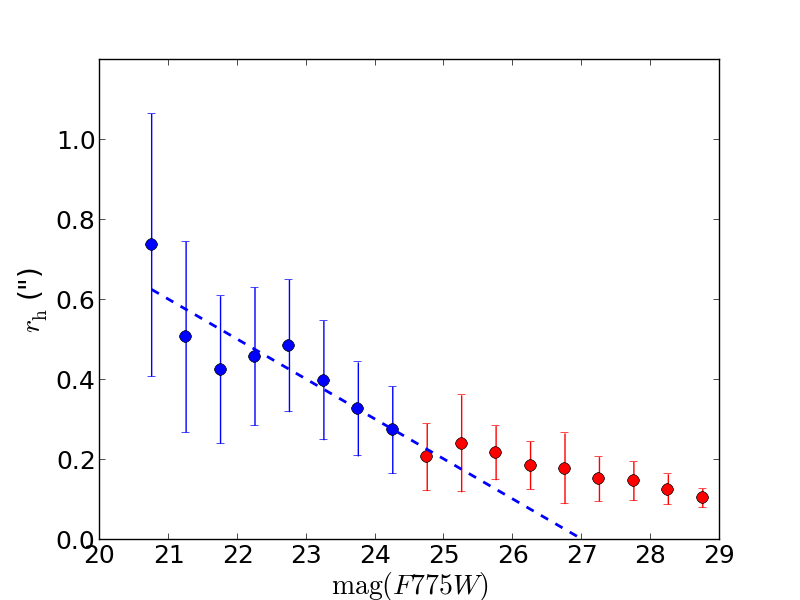}
\caption{Mean PSF-corrected half-light radius vs. magnitude for galaxies measured in the UDF F775W image. Blue dots correspond to galaxies brighter than the VIS limit, and red dots to fainter galaxies within $3\arcsec$ of a bright one. Dots and error bars correspond to the mean and dispersion over all galaxies in the selected magnitude bin. The dashed blue line shows the linear fit to the bright galaxies, highlighting the need for observational data to measure faint galaxy sizes.}
\label{fig:rmag}
\end{figure}

Our galaxy sample is generated based on deep HST images. The UDF survey is one of the very few surveys reaching a magnitude of 29. The magnitude limit for the {\it Euclid} weak-lensing galaxy sample (referred to as the `bright galaxies' in the following) is set to 24.5, which corresponds to a $10\sigma$ detection limit in the VIS instrument \citep{Cropper+16}. Fainter galaxies, up to magnitude 29, are referred to as the `faint galaxies' in our analysis. In Fig.~\ref{fig:rmag} we show the importance of using observed galaxies for the faint population by displaying the average size--magnitude relation of faint galaxies ($24.5<{\rm F775W}$) and comparing it with an extrapolation for the bright galaxies (${\rm F775W} \leq 24.5$). Magnitudes are measured with the {\small MAG\_AUTO} procedure of \texttt{SExtractor} and sizes correspond to PSF-corrected half-light radii measured with the \texttt{SExtractor}/\texttt{PSFEx} method described in the same section below. As already shown in \citet{Hoekstra+17}, the extrapolation from the bright galaxies strongly underestimates the sizes of the faint galaxies.

The downside of using the UDF is its small area ($11.35$ arcmin$^{2}$ after removing saturated stars), which results in a sample-variance issue, since we will simulate thousands of square degrees by sampling from this catalog. The statistics could be increased by using existing or new HST observations, or later the {\it Euclid} deep fields. The latter will however be limited to magnitude 26.5--27. So far only the parallel fields of the HST Frontier Field clusters \citep{Lotz+17} achieve a depth similar to that of the UDF.

The data have been reduced by the UDF team using the \texttt{CALACS} pipeline for the initial calibration and \texttt{MultiDrizzle} \citep{Koekemoer+03} for combining images. All measurements are done on the F775W band, which is included within the VIS filter of the {\it Euclid} survey. We therefore assume that the magnitudes measured in this filter are a good approximation of the VIS magnitudes.

\begin{figure}
\centering
\includegraphics[width=0.5\textwidth,clip]{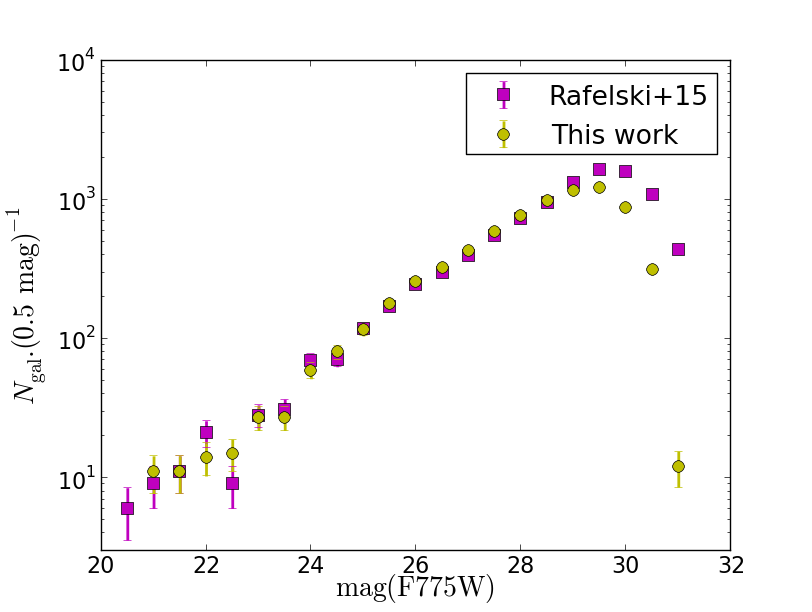}
\caption{Distribution of galaxy magnitudes measured in the UDF F775W image with our detection set up (yellow dots) compared to that of \citet{Rafelski+15} (pink squares), with Poisson error bars.}
\label{fig:raf}
\end{figure}

We measure the galaxy properties using the latest versions of \texttt{SExtractor} (2.31.1) and \texttt{PSFEx} (3.18.0).
We first run \texttt{SExtractor} to generate a catalog of objects that will be used to measure the PSF. Stars are identified in a maximum-surface-brightness-versus-magnitude diagram, and are visually verified on the image. 
Removing saturated objects leaves us with 20 stars in the magnitude range [22, 27]. \texttt{PSFEx} is then run on this star catalog to obtain a PSF model of the UDF survey, which includes spatial variations. The PSF is stable with a variation of less than 1.5\% across the survey area.

\begin{figure}
\centering
\hspace{-1.5cm}
\includegraphics[width=0.5\textwidth,clip]{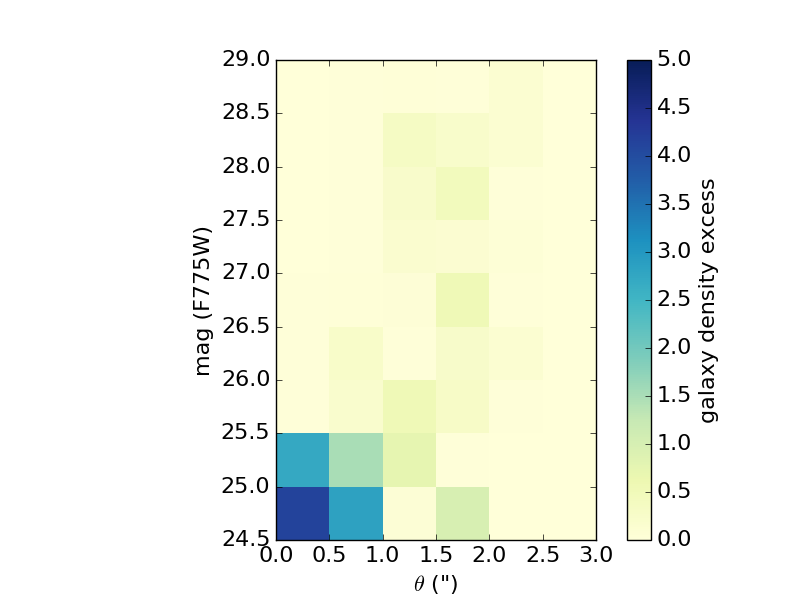}
\caption{Faint galaxy density excess $(N-\bar{N})/\bar{N}$ in the UDF, as a function of magnitude and clustering length. The field density is reached for an excess of zero.}
\label{fig:radpermag}
\end{figure}

We then re-run \texttt{SExtractor} to fit each galaxy with a S\'ersic model convolved with the previously obtained PSF. The parameters we are particularly interested in for generating the simulations are magnitude, half-light radius, S\'ersic index, and ellipticities. In addition, we obtain photometric redshifts by cross-matching our catalog with that of \citet{Rafelski+15}. This is done by selecting the closest match with a maximum separation of 0.3 arcsec (i.e., 10 pixels). Prior to that we verified that the magnitude distributions of the two catalogs match well, which is the case down to magnitude 29, as seen in Fig.~\ref{fig:raf}. \citet{Rafelski+15} detect more objects at magnitudes fainter than 29, but those extra objects include some ambiguous detections. For our simulation purposes, we need high purity and therefore limit our analysis to higher-S/N detections.

To realistically position the faint galaxies, we also measure their clustering around bright ones. We retain only faint galaxies within a separation of $\theta_{\rm lim}$ from a bright galaxy. We choose $\theta_{\rm lim} = 3\arcsec$ for the maximum separation between bright and faint galaxies, which corresponds to about 25~kpc at $z=1$. This choice is justified by Fig.~\ref{fig:radpermag}, which shows the faint galaxy density excess as a function of magnitude and separation. The excess is defined as $(N-\bar{N})/\bar{N}$ where $N$ is the observed galaxy density with clustering and $\bar{N}$ with random positioning, which means that an excess value of zero corresponds to the field density. We see that the clustering is of significant amplitude only for scales lower than $2''\!\!.0$. In addition, galaxies with magnitudes between 24.5 and 25.5 are the most correlated, which is expected since clustered galaxies tend to have similar magnitudes. This also means that the correlations seen in Fig.~\ref{fig:radpermag} are dominated by those between the faintest of the bright galaxies and the brightest of the faint galaxies. In particular, we see no correlations if we make the same plot using only galaxies brighter than magnitude 21 as the bright galaxy sample.

This measured clustering of the faint galaxies depends on the deblending strategy adopted in extracting the catalog from the UDF. An aggressive deblending would allow us to detect more faint blended galaxies (especially in the vicinity of bright ones), but would also misidentify some star-forming regions of bright galaxies as faint objects. On the other hand, a weak deblending would prevent us from detecting most of the faint satellite galaxies. We thus choose a middle-ground deblending strategy using a number of deblending sub-thresholds ({\small DEBLEND\_NTHRESH}) and a minimum contrast parameter for the deblending ({\small DEBLEND\_MINCONT}) of 16 and 0.01, respectively, in \texttt{SExtractor}. In Sect.~\ref{sec:deb} we test the impact of two other deblending schemes on the main results of the paper (one more aggressive and one less aggressive).

The number of galaxies in the bright sample is 244. The number of faint galaxies within $\theta_{\rm lim} = 3 \arcsec$ around bright ones is 189, 333, and 542 for limiting F775W-band magnitudes of 27, 28, and 29, respectively. Although there are more faint galaxies if we account for those without bright neighbors (e.g., 1307 up to $F775W=27$) we use only galaxies within $\theta_{\rm lim}$ to ensure that our faint population reflects that of close neighbors.

\begin{figure*}
\centering
\includegraphics[width=1.0\textwidth,clip]{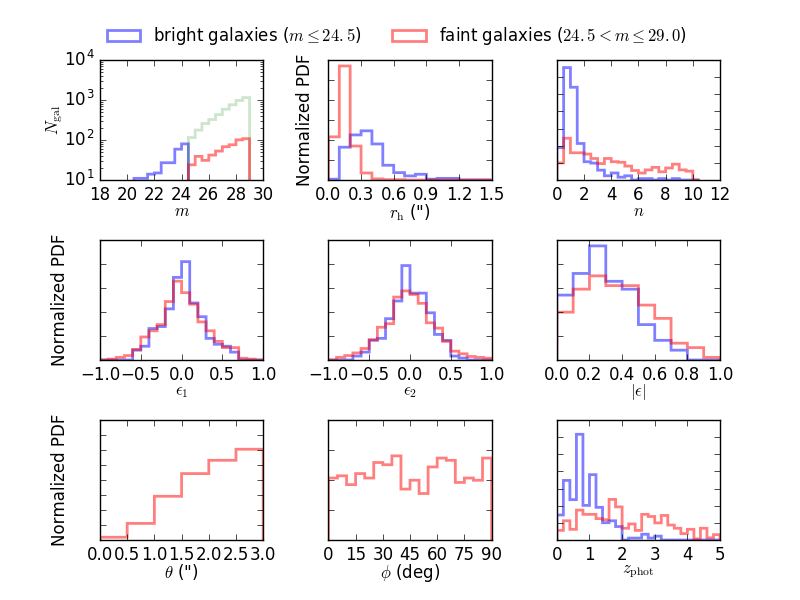}
\caption{Distributions of galaxy parameters measured with \texttt{SExtractor} in the UDF. The panels show histograms of galaxy magnitudes ($m$, {\it top left}), half-light radius ($r_{\rm h}$, {\it top middle}), S\'ersic index ($n$, {\it top right}), ellipticity components ($\epsilon_1, \epsilon_2$, {\it middle left, middle middle}), \aaa{ellipticity modulus ({$|\epsilon|$}, {\it middle right}),} distance to nearest bright galaxy ($\theta$, {\it bottom left}), faint galaxy position angle relative to the nearest bright galaxy major axis ($\phi$, {\it bottom middle})\aaa{, and} photometric redshifts ($z_{\rm phot}$, \aaa{{\it bottom right}}). Blue histograms correspond to bright galaxies ($m \leq 24.5$) and red to faint galaxies ($24.5 < m \leq 29$) lying within $3\arcsec$ of a bright one. The green histogram in the  top left panel shows the magnitude distribution of all faint galaxies up to $m=29$.}
\label{fig:histoobs}
\end{figure*}

Figure~\ref{fig:histoobs} shows the distributions for various measured parameters: magnitude $m$, half-light radius $r_{\rm h}$, S\'ersic index $n$, ellipticity components $\epsilon_1, \epsilon_2$, redshift $z_{\rm phot}$, and distance to the closest bright neighbor $\theta$. In this paper we define the complex ellipticity as $\epsilon = \epsilon_1 + {\rm i}\epsilon_2$, with an absolute value of $|\epsilon|=\left(a-b\right)/\left(a+b\right)$, where $a$ and $b$ correspond to the semi-major and semi-minor axes,
respectively. We also checked whether there is a preferential direction for the clustering by measuring the angle $\phi$ between the position of the faint galaxies relative to their closest bright galaxy center and the semi-major axis of the bright galaxy. We found no significant correlation, perhaps due to the small sample of faint clustered galaxies, and therefore did not test for anisotropic clustering in our simulations. Finally, we measured the correlations between the bright galaxy orientation and that of its faint neighbors and also between the orientations of pairs of faint neighbors belonging to the same bright object patch. We did not find any significant correlation for those quantities (which are not displayed in  Fig.~\ref{fig:histoobs}) and therefore considered the galaxy ellipticity angle as an independent variable. We note however that we considered all faint galaxies together, such that a more refined analysis, which would individually treat faint clustered and unclustered galaxies using their redshift information, could find some correlations in the orientations of the clustered galaxies. This more complex approach would however not qualitatively change our results and is postponed to further studies.

\subsection{Generating galaxy catalogs}

\begin{figure*}
\centering
\includegraphics[width=1.0\textwidth,clip]{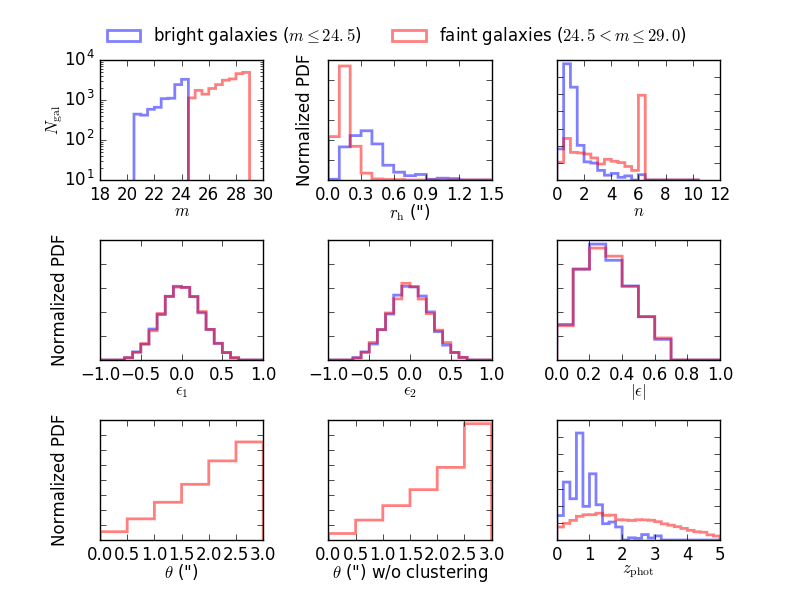}
\caption{Distributions of galaxy parameters generated from our UDF catalog. The panels show histograms of galaxy magnitudes ($m$, {\it top left}), half-light radius ($r_{\rm h}$, {\it top middle}), S\'ersic index ($n$, {\it top right}), ellipticity components ($\epsilon_1, \epsilon_2$, {\it middle left, middle middle}), \aaa{ellipticity modulus ({$|\epsilon|$}, {\it middle right}),} distance to nearest bright galaxy ($\theta$, {\it bottom left}), distance to nearest bright galaxy without clustering ({\it bottom middle})\aaa{, and} photometric redshifts ($z_{\rm phot}$, \aaa{{\it bottom right}}). Blue histograms correspond to bright galaxies ($20.5 \leq m \leq 24.5$) and red to faint galaxies ($24.5 < m \leq 29$) lying within $3\arcsec$ of a bright one.}
\label{fig:histocat}
\end{figure*}

We need to sample from our UDF catalog to simulate a large number of galaxies in order to reach the desired precision on the shear amplitude. These galaxies must all be mutually different and reflect the global properties of the observed population, in particular the covariance between the different parameters. We checked the correlations between parameters, and unsurprisingly found that the half-light radius $r_{\rm h}$, the S\'ersic index $n$, and the clustering (i.e., the number of the faint neighbors $N$ and their separation $\theta$ to the closest bright galaxy) strongly correlate with the magnitude. The different parameters also correlate one with another, in particular $r_{\rm h}$ and $n$, but this degeneracy is broken when splitting the catalog in magnitude bins. Therefore, we construct the conditional probability distribution functions (PDFs) of these parameters given the magnitude bin, from 20.5 to 29 in bins of 0.5: $p(r_{\rm h}|m)$, $p(n|m)$, $p(N|m)$, and $p(\theta|m)$. A magnitude must first be drawn for each object using the magnitude PDF $p(m)$. We recall that in the case of the faint galaxies, only those within $3''$ of a bright galaxy are used to construct the PDFs. Those PDFs are approximated with a trapezoidal function with a bin width chosen so that this model is not significantly different from the full PDF. It is then possible to assign a random value for each parameter, drawing from a uniform distribution between zero and one.

We recall that the magnitude range is [20.5, 24.5] for bright galaxies and \aaa{[}24.5, 29.0] for faint ones. The half-light radius range is \mbox{$0''<r_{\rm h}<1''\!\!.4$} with a bin width of $0''\!\!.1$. For S\'ersic indices, we use 40 different values between zero and ten. We did not use a continuous spectrum of values for the S\'ersic index to speed up the galaxy simulations. Each ellipticity component is drawn from a Gaussian distribution $p(\epsilon_i)$ with a mean of zero and a standard deviation $\sigma_\epsilon=0.26$, which is representative of galaxies of magnitude $24.0 < m < 24.5$ in \citet{Schrabback+18}.

This is how we proceed with the steps in order:
\vspace{0.15cm}

\noindent For each bright galaxy:

1. we draw a magnitude in the range [20.5, 24.5] from $p(m)$;

2. we draw a half-light radius $r_{\rm h}$ and S\'ersic index $n$, sampling from the PDF measured in the galaxy magnitude bin: $p(r_{\rm h}|m)$ and $p(n|m)$;

3. we draw each ellipticity component independently from Gaussian distributions $p(\epsilon_i)$;

4. we determine the number of faint neighbors within $\theta_{\rm lim} =3 \arcsec$ using the PDF corresponding to the bright galaxy magnitude bin $p(N|m)$.
\vspace{0.15cm}

\noindent For each faint galaxy:

5. we draw a magnitude in the range [24.5, m$_{\rm lim}$] from the faint galaxy magnitude PDF $p(m)$;

6. we draw a half-light radius and S\'ersic index, sampling from the corresponding PDFs measured in the galaxy magnitude bin, $p(r_{\rm h}|m)$ and $p(n|m)$;

7. we draw each ellipticity component independently from the same ellipticity distributions as for the bright galaxies, ignoring the small increase in the ellipticity rms observed at fainter magnitudes \aaa{in Fig.~\ref{fig:histoobs} and} in \citet{Schrabback+18};

8. we draw a separation $\theta$ to the closest bright galaxy, sampling from the PDF measured in the faint galaxy magnitude bin $p(\theta|m)$ (Fig.~\ref{fig:radpermag}), with a bin width of $0''\!\!.5$ over the range [$0\arcsec$, $3\arcsec$]. We also draw a random position angle as we found the clustering to be isotropic.

9. Additionally, for each faint galaxy we draw a random position within a $3\arcsec$ circle centered on the bright galaxy to be able to simulate a situation without clustering. In this scheme the number of faint galaxies is the same  with or without clustering. This \corr{assumes} that the change in the galaxy density due to the clustering becomes negligible when approaching the limiting value $\theta_{\rm lim}=3\arcsec$, as can be seen from Fig.~\ref{fig:radpermag}.\\

The limiting magnitude m$_{\rm lim}$ for the faint galaxies takes different values, chosen to check the depth at which the faint galaxies need to be included in the simulations in order not to bias the shape measurement of the bright ones.

We use single S\'ersic profiles for galaxy shapes. Although this is a simplistic model, it is computationally fast and allows us to use the same model for bright and faint galaxies. Other more sophisticated models, such as combining two S\'ersic profiles to account for the bulge and disk, should however not qualitatively change the results of the paper, although the quantitative effect of the faint galaxies might vary for different galaxy populations.

Finally, we need to apply some corrections to the sampled catalog so that every galaxy can be properly simulated. We require that the ellipticity modulus be lower than 0.7, redrawing both ellipticity components for objects that do not satisfy this criterion. Larger ellipticities lead to some unrealistic, truncated galaxy profiles as the semi-major axis of the largest galaxies can reach the edge of the galaxy patch\aaa{, which is of $64\times64$ pixels with a truncation radius fixed at $4.5~r_{\rm h}$}. We could also avoid this by increasing the patch size or by using more complex models such as a bulge and disk decomposition, but these approaches would be computationally more demanding. The fraction of galaxies for which we need to redraw ellipticities is below 3\%, so this criterion should not qualitatively change our results. Furthermore, \texttt{GalSim} cannot simulate galaxies with a S\'ersic index out of the range [0.3, 6.2]. Each galaxy with $n$ outside this range has its S\'ersic index set to the acceptable limit. This explains the peak at $n=6.2$ in the distribution of the S\'ersic index of the faint galaxy population in Fig.~\ref{fig:histocat}. We could also have chosen to cut out those galaxies, but this would more strongly distort the S\'ersic index PDF, or to distribute them on a range of values; the latter approach would however require us to make some assumptions on the S\'ersic index measurement errors. We note that all bright galaxies have $n$ within the cited range, which is not the case for the faint galaxies, especially at very faint magnitudes where $n$ has an almost uniform probability distribution between zero and ten, as measured with \texttt{SExtractor} (see Fig.~\ref{fig:histoobs}). This is because these galaxies cover only a few pixels, on which one cannot reliably fit a S\'ersic profile. The size and magnitude of these objects however remain accurate since they do not require us to measure the surface brightness profile. This limitation only concerns the faintest galaxies, and is less problematic, since we are not trying to measure the shape of these galaxies. In addition, their fluxes and sizes should be sufficient to assess the correlated noise due to the extension of these objects on a few pixels in the sky background. Except for the few corrections mentioned above, we see good agreement between the observed catalog (Fig.~\ref{fig:histoobs}) and the sampled one (Fig.~\ref{fig:histocat}).

We see in Figs.~\ref{fig:radpermag} and \ref{fig:histoobs} that there are almost no observed galaxies below $0''\!\!.5$ for magnitudes fainter than 25.5. This is not a physical property of the galaxy population, but shows limitations in the clustering measurement around bright galaxies. The mean size of the bright galaxies is $r_{\rm h}=0''\!\!.38,$ meaning that they are masking faint galaxies in their close vicinity. We fit a power law to the closest neighbor--distance distribution to extrapolate the galaxy clustering into the inner $0''\!\!.5$ radius for each magnitude bin. In Fig.~\ref{fig:histocat} we see that the clustering extends to this first bin. Finally, the choice of the deblending strategy when detecting objects in the data may have a significant influence on the clustering at the smallest scales. This effect is discussed in Sect.~\ref{sec:deb} where we test the impact of several deblending setups for the sensitivity of shear measurements on the clustering of the faint population.

\subsection{Simulating galaxies}

\begin{figure*}
\centering
\includegraphics[width=0.4\textwidth,clip]{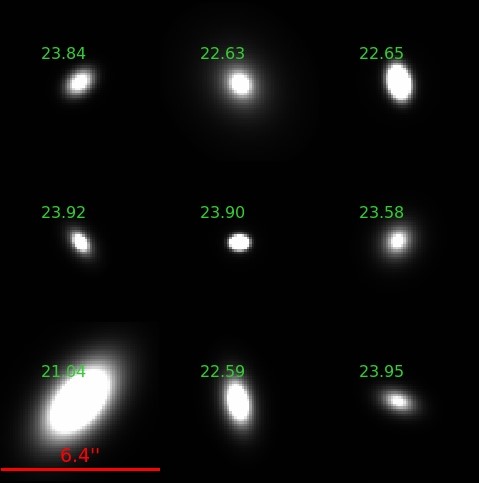}
\includegraphics[width=0.4\textwidth,clip]{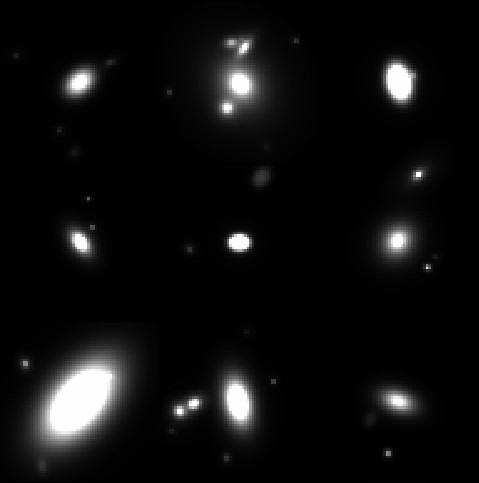}

\includegraphics[width=0.4\textwidth,clip]{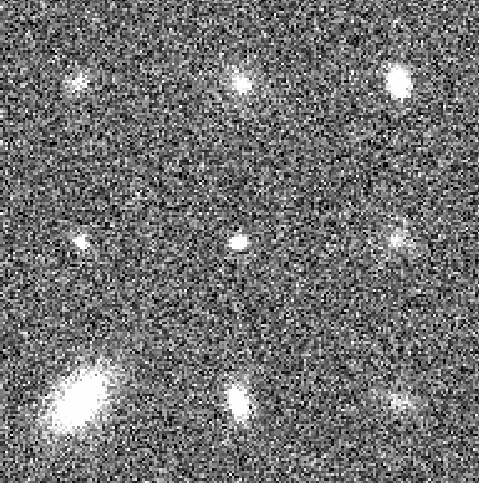}
\includegraphics[width=0.4\textwidth,clip]{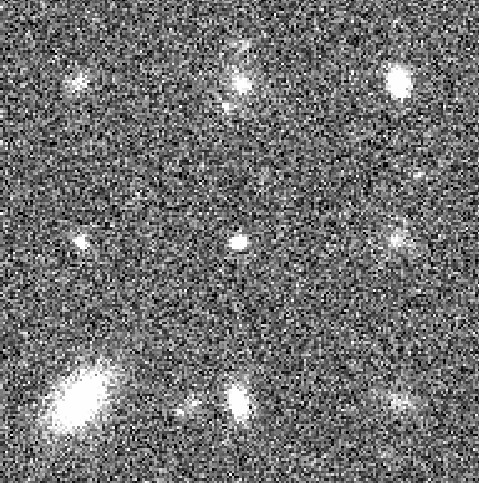}
\caption{Image simulations, with bright galaxies on a grid ({\it left}), and with the faint galaxies  down to magnitude 29 added, including clustering properties ({\it right}). The {\it upper panel} shows noiseless simulations and the {\it bottom} one simulations with realistic Gaussian noise. This sub-image presents nine tiles of $6''\!\!.4\times6''\!\!.4$ each. The scale is given by the red line in the upper left panel. The numbers in the same panel correspond to the magnitudes of the bright galaxies. The two right panels are populated with an identical set of 30 faint galaxies.}
\label{fig:simsim}
\end{figure*}

\begin{figure*}
\centering
\includegraphics[width=1.0\textwidth,clip]{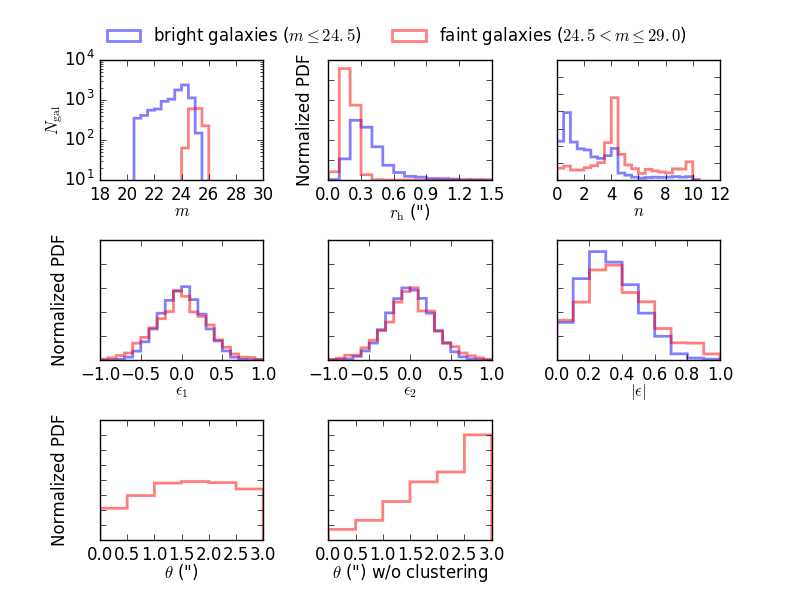}
\caption{Distributions of galaxy parameters measured in our VIS-like noisy image simulations. The panels show histograms of galaxy magnitudes ($m$, {\it top left}), half-light radius ($r_{\rm h}$, {\it top middle}), S\'ersic index ($n$, {\it top right}), ellipticity components ($\epsilon_1, \epsilon_2$, {\it middle left, middle middle}), \aaa{ellipticity modulus ({$|\epsilon|$}, {\it middle right}),} distance to nearest bright galaxy ($\theta$, {\it bottom left}), and distance to nearest bright galaxy without clustering ({\it bottom middle}). Blue histograms correspond to bright galaxies and red to faint galaxies lying within $3\arcsec$ of a bright one. For this figure, bright galaxies were measured in simulations including only galaxies with $20.5\leq m\leq24.5$, and faint galaxies were measured in simulations including only galaxies with $24.5<m\leq29$ (see text for details).}
\label{fig:histosim}
\end{figure*}

Galaxy images are simulated via the \texttt{GalSim} software \citep{Rowe+15}, with properties from the input catalog generated in the previous section. For each galaxy, we first draw a S\'ersic surface brightness profile using the `galsim.Sersic' function with $n$, $r_{\rm h}$, and $m$. We then add ellipticity from the input $\epsilon_1$ and $\epsilon_2$ using the \texttt{GalSim} function `galsim.Shear(${\rm g1} =\epsilon_1$, ${\rm g2} =\epsilon_2$)' with keywords ${\rm g1}$, ${\rm g2}$ corresponding to our ellipticity definition in \texttt{GalSim}. Finally we add a fixed shear value ($\gamma_1^{\rm t}$, $\gamma_2^{\rm t}$) with the same function. The choice of these values is discussed in Sect.~\ref{subsec:biasdef}. For simulations containing faint galaxies, we recompute the faint galaxy positions by applying the lensing transformation due to the input shear value and centered on the closest bright neighbor. This is to preserve a realistic positioning of bright and faint galaxies relative to each other in the lens plane, although we found that this has a negligible effect on the measured shear bias. \aaa{We also apply the same input shear to the faint galaxies as that applied to the bright ones. The impact of this choice is discussed in Sect.~\ref{subsec:bias}.}

Galaxy images are then convolved with the PSF before being added to the image. The PSF is the average of three symmetric Airy PSFs for a 1.2 m diameter telescope with an obscuration of 0.3 m, computed at wavelengths of 600, 700, and 800~nm. Although a single wavelength is already a good approximation of the expected {\it Euclid} PSF \citep{Hoekstra+17}, adding the extra wavelengths allows us to better represent the large passband of the VIS filter. We also assume a flat spectral energy distribution with no spatial dependence for every galaxy, such that the three components of the PSF are equally important for each object. In this paper we do not assess the effect of PSF anisotropy or variability on the shape measurement, meaning that a simple model for the PSF is sufficient and saves computational time.

Bright galaxies are positioned onto a grid and separated from each other by $6''\!\!.4$. We choose this value so as to include the clustering measured in a $3\arcsec$ radius, and so that the galaxy patch is $64\times64$ pixels which speeds up computation. We use a grid instead of random positions to avoid any contamination from bright galaxy blending. The faint galaxies are positioned around bright galaxies according to the observed clustering in terms of numbers and separation from the bright galaxies. As we did not find any evidence for anisotropic clustering, we place the faint galaxies at random angles. All galaxies are shifted by a random subpixel value to avoid
always being centered on the middle of a pixel.

Each image encompasses 10\,000 bright galaxies, plus the additional faint galaxies for half of the simulations, and mimics VIS images. In particular, the pixel size is $\ell=0''\!\!.1$ \citep{Laureijs+11} and the exposure time $t_{\rm exp}$ corresponds to a co-addition of three single exposures of $565~{\rm s}$ each \citep{Laureijs+11}. In this study we ignore the complication arising from half the data being planned to have a fourth exposure. The CCD gain is set to $g=3.1~{\rm electrons/ADU}$ \citep{Niemi+15}. Galaxy fluxes $F$ are defined in \corr{analog-to-digital units (ADU)} according to the following equation:
\begin{equation}
\label{eq:flux}
F^{\rm ADU} = \frac{t_{\rm exp}}{g} 10^{-\left(m - {\rm ZP}\right)/2.5}\;,
\end{equation}
where ${\rm ZP}$ is the instrumental magnitude zero-point adjusted to reflect the S/N of {\it Euclid} galaxies as discussed in detail below.

Once every galaxy is simulated we add Gaussian noise to the image. In this approach we neglect the Poisson noise term from the photon counts of the sources. This overestimates the S/N of the brightest galaxies but saves computational time and allows us to implement a background-noise cancellation, which we introduce in Sect.~\ref{subsec:ngal}. We note that this simplification does however not affect the S/N of the faint galaxies, the impact of which is the main interest of this study. As we place galaxies on a grid, we cannot estimate the rms background by matching the source galaxy density to that expected for {\it Euclid}, as done by \citet{Hoekstra+17}. Instead we follow the approach of \citet{Tewes+19}. We set the read-out noise level to $\sigma_{\rm readout}=4.2~{\rm e^{-}}$ \citep{Cropper+16} and the sky background to $\mu_{\rm sky}=22.35~{\rm mag}\,{\rm arcsec}^{-2}$ \citep{Refregier+10} and compute the noise rms assuming Poisson errors on the number of electrons measured due to the background,
\begin{equation}
\label{eq:n1}
F_{\rm sky}^{{\rm e^{-}/pixel}} = \ell^2 t_{\rm exp} 10^{-\left(\mu_{\rm sky} - {\rm ZP}\right)/2.5}\;,
\end{equation}
\begin{equation}
\label{eq:n2}
\sigma_{\rm bkg}^{{\rm e^{-}/pixel}} = \sqrt{F_{\rm sky}^{{\rm e^{-}/pixel}} + \sigma_{\rm readout}^2}\;.
\end{equation}
The noise rms is then converted into ADU per pixel by dividing by the gain. We adjust the zero-point so that a galaxy of $m=24.5$ has a S/N=10 on average, as required for the {\it Euclid} survey \citep{Cropper+16}. The S/N is estimated as the ratio between {\small FLUX\_AUTO} and {\small FLUXERR\_AUTO}, as measured by \texttt{SExtractor}. This leads to an instrumental zero-point of \mbox{${\rm ZP}=24.0$} and a noise of $\sigma_{\rm bkg}=3.15~{\rm ADU/pixel}$. The image zero-point is higher than the instrumental one by $2.5 \log{(t_{\rm exp}/g)}$ to account for the image exposure time and gain. The image zero-point is equal to $30.84$. We recall that the magnitudes we use are measured in the HST F775W filter, which is included within the VIS filter. Our simulated galaxy magnitudes are therefore an approximation to the VIS ones, with realistic PSF and noise. In particular, we neglect the bluer contribution of the VIS filter to the galaxy magnitudes, although it is included in the PSF. We refer to our simulated magnitudes as $m$ in the rest of the paper. We use the same random seed to determine the noise in both images: the image with only the bright galaxies and that with the bright and faint galaxies, applying the exact same noise contribution to both.

An example of a sub-image is shown in Fig.~\ref{fig:simsim}; in the left panels we show a few simulated bright galaxies; and in the right panels we add the faint galaxies up to $m=29$ to the image. The top panels are without noise, while the bottom panels have the noise added to the image. We immediately see that the faintest galaxies become buried in the noise and will no longer be detected, but will contribute to the surface brightness around the target source.

Using the same \texttt{PSFEx} and \texttt{SExtractor} procedure as for the UDF data, but with the PSF of our VIS-like images, we measure the properties of the simulated galaxies to compare them with the input of the simulation. To clearly distinguish bright from faint galaxies, we take the measurements separately in simulations including only one of the two populations. This is the only simulation in the paper to be run without including the bright galaxies. The histograms for the magnitude, S\'ersic index, half-light radius, ellipticities, and closest bright neighbor distance are shown in Fig.~\ref{fig:histosim}. We see the same problem as before for the S\'ersic indices of the faint galaxies, which cannot be reliably measured. The most striking point when comparing to the input catalog (Fig.~\ref{fig:histocat}) however is the disappearance of a large number of faint galaxies, with an accompanying distortion of the clustering distance distribution. This occurs because most faint galaxies are no longer detected, since we added a realistic {\it Euclid} noise level. We note that a small number of faint galaxies are detected and could therefore be accounted for when measuring the shear. We treat them as undetected faint objects in the measurement pipeline however, since the mitigation of the impact of the faint galaxies is beyond the scope of this paper. When we include the clustering in the simulation procedure, faint galaxies appear more clustered than in the original catalog; this is because we can detect only the brightest of the faint galaxies, which are the most clustered. Finally, we note that the ellipticity distributions are almost unchanged from Fig.~\ref{fig:histocat} to Fig.~\ref{fig:histosim}. This is because no shear has yet been applied and we detect only galaxies with a high S/N, for which the ellipticity is not strongly affected by the noise. The effect of the noise can nonetheless be seen in the tails of the ellipticity distributions that are slightly wider in Fig.~\ref{fig:histosim}, especially for the faint galaxies.

\section{Shape-measurement algorithms}
\label{sec:shape}

Each shape-measurement algorithm responds differently to noise issues \citep[see e.g., the {\small GREAT3} challenge; ][]{Mandelbaum+15}. Correlated noise induced by the faint galaxies might therefore affect each algorithm differently. In order to obtain a comprehensive overview of this effect we select three shape-measurement algorithms that are representative of the three main types of existing methods. We use \texttt{SExtractor}/\texttt{PSFEx} as our model-fitting technique, \texttt{MomentsML} \citep{Tewes+19} as a machine-learning algorithm, and a moment-based \texttt{KSB+} algorithm developed in \citet{Erben+01}. \texttt{SExtractor}/\texttt{PSFEx} and an earlier version of \texttt{MomentsML} \citep[named \texttt{MegaLUT}:][]{Tewes+12} were respectively ranked second and fourth in the {\small GREAT3} challenge, and therefore represent some of the best contemporary shear measurement methods, while \texttt{KSB+} is more classical and computationally inexpensive. It is important to note that we do not try to optimize these algorithms to mitigate the impact of the faint galaxies, as our goal is to quantify the impact of neglecting them in the calibration simulations. The different algorithms are however optimized to have low multiplicative and additive biases in the simulations that have the bright galaxies only.

It is also important to note that the estimation of the background can have a significant impact on the shear biases \citep[see][for an example of KSB measurements]{Hoekstra+17}. We therefore apply the same background estimate for all three methods so that we can consistently compare the three algorithms. The standard estimate in our analysis is the one from \texttt{SExtractor}, which is computed at the galaxy-patch level ({\small BACK\_SIZE}$=$64, {\small BACK\_FILTERSIZE}$=$3). All three measurement algorithms are then applied on the background-subtracted images. We note that by construction the mean background is equal to zero in our simulations when the faint galaxies are not included. We further check the impact of background estimates in Sect.~\ref{sec:bk}, by measuring galaxy shapes without subtracting the background.

\aaa{The initial detection of objects is done with \texttt{SExtractor} for every algorithm.}

\subsection{\texttt{SExtractor}/\texttt{PSFEx}}

\aaa{Although \texttt{SExtractor}/\texttt{PSFEx} is routinely used for object detection, a model-fitting shape-measurement implementation has been developed in} versions 3.18.0 and 2.31.1 of these software packages. \texttt{PSFEx} measures the PSF properties using stars. This model is then convolved with a surface brightness profile and fitted to galaxies with \texttt{SExtractor}.

We estimate the PSF from 10\,000 stars simulated in the same way as we simulate galaxies. These stars have magnitudes in the range [20.5, 24.5] and we also apply a subpixel random position shift. We do not add noise to these images since this paper does not probe the quality of the PSF reconstructions, but assumes instead that the PSF is perfectly known for each galaxy. The configuration of the algorithm is similar to what is described in Appendix C1 of \citet{Mandelbaum+15}, which describes the results of the {\small GREAT3} challenge. In particular, we allow for a subpixel sampling of the PSF with a subpixel size of 0.4 pixels, in contrast to 0.3 pixels for {\small GREAT3}. The size of the patch on which the PSF is modeled is $40\times40$ subpixels, which corresponds to more than ten times the expected {\it Euclid} VIS PSF full width at half maximum \citep[$0''\!\!.17$ according to ][]{Cropper+16}. These choices are found to be a good trade-off between performance and computational time.

Galaxies are fitted with a single S\'ersic profile, in which the centroid position, amplitude, effective radius, axis ratio, position angle, and S\'ersic index are free parameters. The fit is performed using the Levenberg-Marquardt algorithm.

We also apply an inverse-variance weighting scheme to each ellipticity component of every galaxy:
\begin{equation}
\label{eq:wsex}
w_i = \frac{1}{\sigma_i^2+\sigma_\epsilon^2}\;,
\end{equation}
where $\sigma_i$ is the error on the measurement of component $i$ of the ellipticity and $\sigma_\epsilon=0.26$ is the shape noise per ellipticity component.

\subsection{\texttt{MomentsML}}

Shear measurements labeled `\texttt{MomentsML}' are obtained with a supervised machine-learning method presented in \citet{Tewes+19}. 
The algorithm uses galaxy shape parameters computed from adaptive moments of the observed images as input features to the machine learning. 
Based on these features, a group of artificial neural networks then regresses a shear estimate for each galaxy. 
In particular, the algorithm predicts point estimates and weights for each component of the shear, with the setup described in Sect. 8 of \citet{Tewes+19}. 

Before applying the method to a data set, the networks are trained on image simulations of the forward observing process. 
A key aspect of this training is the propagation of many {\it realizations} of each observation through the networks.
The optimization of the network parameters aims at obtaining estimates that are statistically accurate over these ensembles of realizations.
Through this mechanism, the machine learning is made aware of the noisiness in the input features (both photon noise and pixellation), which would otherwise lead to biases.
 
For the sensitivity study conducted in this paper, we deliberately train the method using only clean single S\'ersic-profile galaxies, without blends or contamination by other sources.
Also, the input features are computed from moments measured with simple elliptical Gaussian weighting functions, as discussed in \citet{Tewes+19}.
No masking or segmentation of the galaxy images is performed.

\subsection{\texttt{KSB+}}

The \texttt{KSB+} formalism computes PSF-corrected galaxy ellipticity estimates from measurements of galaxy and stellar weighted brightness moments \citep{KSB95,Luppino+97,Hoekstra+98}. For our analysis we employ the \citet{Erben+01} implementation of the \texttt{KSB+} algorithm as further detailed in \citet{Schrabback+07,Schrabback+10}. We use the same sample of 10\,000 point-like sources as for the \texttt{SExtractor} and \texttt{PSFEx} method to measure the moments of the PSF.

For our current analysis we decided to not include the correction for noise-related multiplicative biases derived by \citet{Schrabback+10}, mainly because of differences in the characteristics of our simulations and the STEP2 simulations \citep{Massey+07} employed to compute this correction. Since this correction would be the same in both the case with and without the faint galaxies, it is not a concern for our analysis; we are primarily interested in the relative change of the bias due to the inclusion of the faint galaxies in the simulations, and not in the absolute value of the bias.

Following \citet{Schrabback+18} we compute the dispersion of the noisy ellipticity estimates in magnitude bins and define shape weights $w(m)=1/\sigma_\epsilon^2(m)$ via the magnitude-interpolated dispersion $\sigma_\epsilon(m)$.

\section{Shear bias measurement}
\label{sec:bias}

\subsection{Bias definition and shear input values}
\label{subsec:biasdef}

We estimate the bias in the shear by comparing the measured shear values $\gamma_i$ to the input true shears $\gamma_i^{\rm t}$. The index $i$ refers to the two components of the shear. We model the bias as a linear function of the true shear:
\begin{equation}
\label{eq:bias}
\gamma_i-\gamma_i^{\rm t}=\mu \; \gamma_i^{\rm t}+c_i\;,
\end{equation}
\noindent where $\mu$ is the multiplicative bias and $c_i$ the additive bias.

As in \citet{Hoekstra+17}, we assume that the multiplicative bias is the same for both components of the shear. We set $\gamma_1^{\rm t}=0$ and $\gamma_2^{\rm t}$ to $101$ different values between $-0.06$ and $0.06$, with a step of $0.0012$. We also verified that fixing $\gamma_2^{\rm t}$ to zero and varying $\gamma_1^{\rm t}$ gives similar results on our fiducial simulation set.

\subsection{Number of galaxies}
\label{subsec:ngal}

To achieve the statistical precision on the cosmological parameters probed, {\it Euclid} will require the combined systematic biases on shear measurement to be lower than $\mu<2\times10^{-3}$ and $c<10^{-4}$. The residual uncertainty is set by the precision with which the bias is determined in the simulations. We then want to probe the variation in these parameters with a precision at least an order of magnitude lower, that is, $\delta\mu<2\times10^{-4}$ and $\delta c<10^{-5}$. We find no strong variation in $c$, and therefore we concentrate on $\mu$ in the rest of the paper. \aaa{The lower impact on $c$ could be explained by the fact that faint galaxies} 
are placed randomly or following an isotropic clustering around bright galaxies. 
We furthermore use a constant PSF with circular symmetry such that no additive bias is introduced at the PSF level. In principle the number of galaxies $N_{\rm gal}$ required to reach a precision $\delta\mu$ is given by \citep[e.g.,][]{FenechConti+17}:
\begin{equation}
N_{\rm gal} = \left(\frac{\sigma_\epsilon}{\delta\mu~|\gamma|}\right)^2,
\end{equation}
\noindent where $\sigma_\epsilon=0.26$ is the dispersion of galaxy ellipticities and $|\gamma|$ is the shear modulus applied in the simulations. For a shear modulus of $0.03$ on average, we need $1.9\times10^{9}$ galaxies. This number can however be reduced through noise cancelation. We use both shape-noise cancelation \citep{Massey+07} and background-noise cancelation.

We simulate the same galaxy with different rotation angles, chosen so that the mean intrinsic ellipticity over all angles is equal to zero, and keeping all other parameters fixed. This significantly reduces the noise due to the intrinsic ellipticities. We use two rotation angles: $0$ and $90$ degrees. We also tried four rotation angles, as done in \citet{FenechConti+17}, but found that with our setup these extra two rotations ($45$ and $135$ degrees) improve the precision on $\delta\mu$ by a factor smaller than \smash{$\sqrt{2}$} and are therefore inefficient. In the rotated images, the faint galaxies are also rotated along with their positions. This is to keep the same pattern between bright and faint galaxies and only cancel the shape noise due to the bright galaxies. If we were to not do this a faint galaxy close to a bright galaxy minor axis would end up along the major axis in the rotated frame, which is not desirable.

The use of Gaussian noise, although less realistic than Poissonian, allows us to reduce the impact of the background noise. We build two identical images, one where the Gaussian noise is added and a second one where the same noise realization is subtracted. Therefore if a galaxy appears stretched due to a bright noise pixel, it will be shortened along the same direction in the image where the noise is subtracted. Taking the average ellipticity measured on these different images further increases the precision on $\delta\mu$. The improvement depends on each measurement method but is significant in all three cases, and can reach values of up to four times better than without the background noise correction. We also note that this trick is computationally very fast, as we only need to add different noise, and do not have to draw galaxies again, which is the slowest step in our simulation pipeline. Finally, on our final set of simulations 
we verified that this technique does not distort the average shear estimates, but only improves the errors on the measured biases.

In our final simulation design we create images with 10\,000 bright galaxies. Each image is simulated with two galaxy rotation angles ($0$ and $90$ degrees) and two noise realizations (adding and subtracting Gaussian noise). Each galaxy is therefore simulated four times and the shear measurement obtained for this galaxy is the average of the ellipticities measured on those four images. We do the same for the second set of simulations, which contain the same bright galaxies and also the faint ones. Applying these noise corrections and using the sampling of the input shear values described in Sect.~\ref{subsec:biasdef}, we find that an approximate number of $8\times10^{7}$ galaxies (including the two rotations and the added and subtracted Gaussian noise) is sufficient to reach an accuracy better than $\delta\mu=2\times 10^{-4}$. This is more than an order of magnitude smaller than the number of galaxies required without noise cancelation. We also note that new techniques are being developed to avoid shape-noise cancelation by measuring the shear response of individual galaxies \citep{Pujol+19}. Although it seems to be a promising way of decreasing the required number of simulated galaxies to reach a given shear accuracy, we do not explore this method here.

\section{Effect of unclustered faint galaxies}
\label{sec:unclus}

We start by considering the effect of unclustered faint galaxies following \citet{Hoekstra+17}, and explore the impact of clustering in the following section.

\subsection{Bias for an unclustered faint population}
\label{subsec:bias}

\begin{figure}
\includegraphics[width=0.5\textwidth,clip]{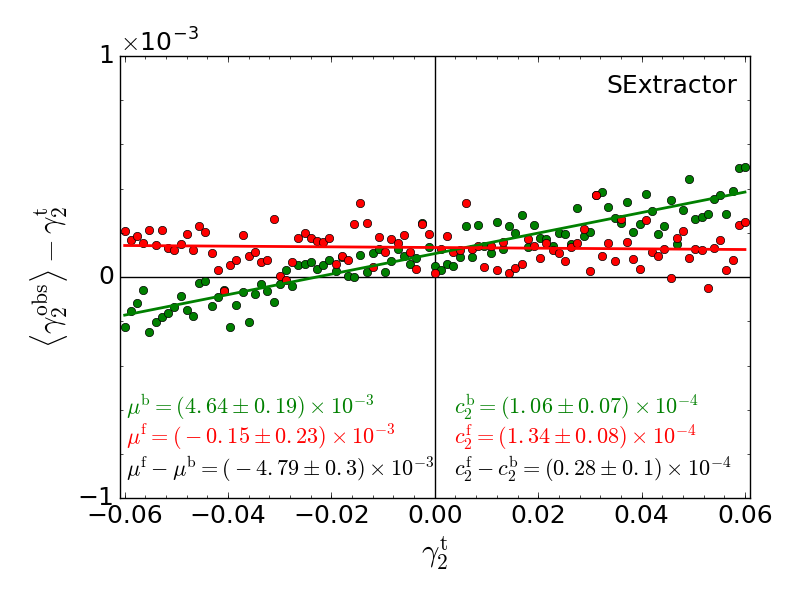}
\includegraphics[width=0.5\textwidth,clip]{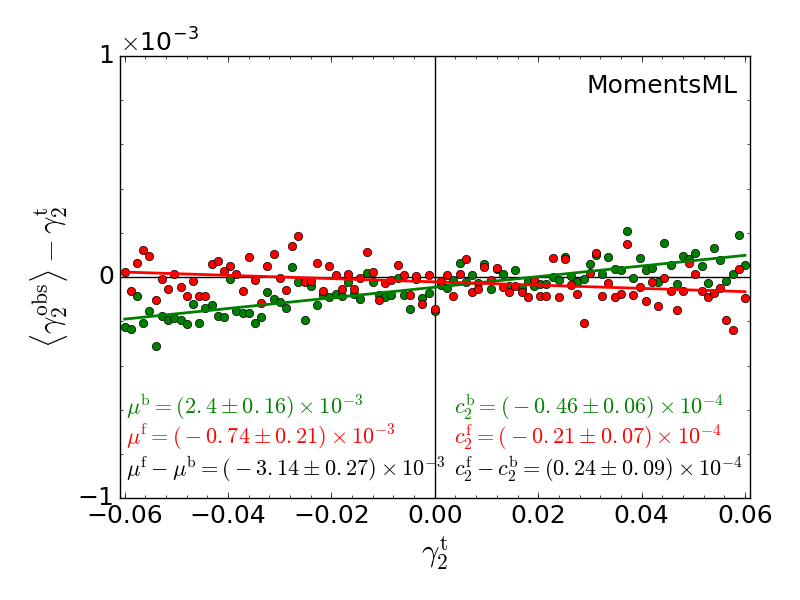}
\includegraphics[width=0.5\textwidth,clip]{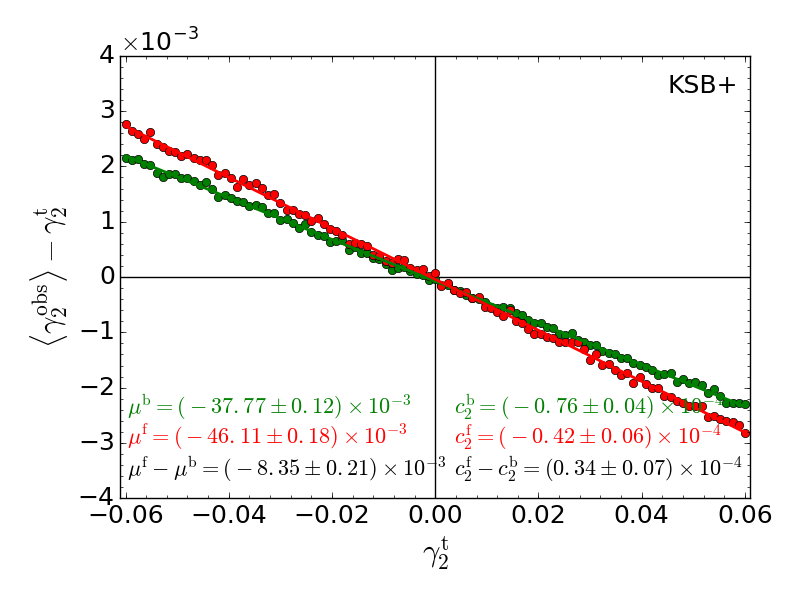}
\caption{Measurement of shear bias from about 20 million noise-canceled shear estimates. {\it Top}: \texttt{SExtractor}/\texttt{PSFEx} measurement. {\it Middle}: \texttt{MomentsML}. {\it Bottom}: \texttt{KSB+}. Green dots represent shear values measured on the bright galaxies of the simulations including only the bright galaxies, and red dots correspond to the values measured on the bright galaxies of the simulations including both bright and faint galaxies up to $m=29$. Multiplicative and additive biases are displayed with the same color code and the difference between the two sets of simulations is shown in black.}
\label{fig:mu}
\end{figure}

We measure the multiplicative bias by fitting a linear relation between the measured and true shear described in Eq.~(\ref{eq:bias}), leaving $\mu$ and $c$  free to vary. Examples of these relations are shown in Fig.~\ref{fig:mu} for the three different algorithms. Each point in this plot corresponds to the average ellipticity over 800\,000 bright galaxies, among which 200\,000 are individual objects and the other 600\,000 correspond to the extra realizations of shape noise and background noise. Since there are 101 different input shear values, the number of shear measurements (i.e., not counting the different angle and noise realizations of the same galaxy) used in the estimation of the biases is about 20 million. We also display the measurements when the faint galaxies with magnitude $24.5<m\leq 29$ are included. We note that in this second case only the shapes of the bright galaxies are measured even if some faint galaxies can be detected in the image. 

\aaa{This is to avoid any leakage of the faint galaxy sample into the bright population, as \citet{Samuroff+18} found that it can introduce a multiplicative bias of the order of 1\%. For comparison we quantify this effect by measuring the mean true ellipticity of all detected objects in the two sets of simulations. We find an increase of about 1\% in the number of objects detected when including the faint galaxies and a difference of the order of $10^{-4}$ in the mean ellipticity which would translate into a multiplicative bias of the same order as that reported in \citet{Samuroff+18}. This highlights the importance of selection bias for weak-lensing analysis. By using only the bright galaxies detected in both simulations with and without the fainter galaxies, we prevent any selection bias in this study, allowing us to better isolate the impact of the undetected objects which is a separate issue from that of the selection function.}

The three methods perform differently. The multiplicative biases achieved for the bright galaxies only are $\mu^{\rm SEx}=(4.64\pm0.19) \times 10^{-3}$, $\mu^{\rm ML}=(2.40\pm0.16) \times 10^{-3}$, and $\mu^{\rm KSB}=(-37.77\pm0.12) \times 10^{-3}$ for \texttt{SExtractor}/\texttt{PSFEx}, \texttt{MomentsML}, and \texttt{KSB+}, respectively. Although the goal of the paper is not to compare the different measurement methods, we note that more refined techniques, such as model fitting and machine learning, are less biased than a simple KSB approach. The lower accuracy of \texttt{KSB+} could be related to the lack of noise-bias correction in our implementation of the \texttt{KSB+} algorithm. Also, the good accuracy of the \texttt{SExtractor} measurements could be linked to the fact that the galaxies used in our simulations have their properties measured on the UDF images with the same software. We note also that the accuracy of the \texttt{SExtractor}/\texttt{PSFEx} measurements is better than that expected from noise bias \aaa{\citep[e.g.,][]{Kacprzak+12,Refregier+12}}, suggesting some fortuitous cancelation of biases with the parameters chosen while optimizing the method on the bright-galaxy case. These achievements are nonetheless promising for meeting the requirements of the {\it Euclid} survey.

The different algorithms also show different sensitivity to the faint galaxy noise, as indicated by the values of $\Delta\mu = \mu^{\rm f} - \mu^{\rm b}$. When including faint galaxies up to $m=29$, we find $\Delta\mu^{\rm SEx}=(-4.79\pm0.30) \times 10^{-3}$, $\Delta\mu^{\rm ML}=(-3.14\pm0.27) \times 10^{-3}$, and $\Delta\mu^{\rm KSB}=(-8.35\pm0.21) \times 10^{-3}$. The error on $\Delta\mu$ is calculated as the quadratic sum of the errors on $\mu^{\rm f}$ and $\mu^{\rm b}$. In this calculation we neglect the correlations between $\mu^{\rm f}$ and $\mu^{\rm b}$, such that the precision on $\Delta\mu$ might actually be better than that of our conservative approach. We see that \texttt{MomentsML} is the least affected by the faint galaxies, followed by \texttt{SExtractor} and then \texttt{KSB+}, but all three methods present significant shifts in their multiplicative bias due to the unresolved galaxies, at the level of a few times $10^{-3}$. We also note that $\mu$ becomes more negative when including the faint galaxies. The faint galaxies tend to distort the bright galaxy shapes in a direction that is uncorrelated with the input shear. On average, this will lower the amplitude of the shear estimates, characterized by ($1+\mu$). 

\aaa{Our choice of applying the same input shear to the faint and bright galaxies however correlates the distortions due to the faint galaxies with the input shear. This tends to increase the amplitude of the shear compared to a case where the observed ellipticities of the faint objects are completely independent from the input shear. Quantitatively, we find $\Delta\mu^{\rm SEx}=(-6.88\pm0.28) \times 10^{-3}$, $\Delta\mu^{\rm ML}=(-4.29\pm0.26) \times 10^{-3}$, and $\Delta\mu^{\rm KSB}=(-11.68\pm0.22) \times 10^{-3}$ for \texttt{SExtractor}, \texttt{MomentsML}, and \texttt{KSB+,} respectively, when not shearing the faint galaxies. Since the faint galaxies are typically at the same or a larger redshift than their bright neighbors, they are likely affected by the same foreground structures, plus extra contributions from more distant ones. The correlations between the direction of the effect of the faint galaxies and the input shear should therefore lie between the two reported cases: largest correlations with identical shear, and lowest correlations without shearing the faint galaxies. We use the former case since it provides us with a conservative estimate of $\Delta\mu$, in the sense that the absolute value of the bias will be larger with the exact shearing of the faint galaxies.}

We note the presence of a significant additive bias (up to $10^{-4}$) with all three methods. Although this value is small, it is puzzling since we use a purely round PSF in our simulations. We conducted several tests to try to understand this bias, and excluded the possibility that it comes from the positioning of galaxies on a grid, from the subpixel shift of galaxy centers, or from the shape- or background-noise cancelation. To assess whether this bias is due to the simulations or the shape-measurement algorithms, we compared the ellipticity measurements in images generated for null shear and in the same images rotated by 90 degrees. The noise should be exactly the same in the two images and for a galaxy with ellipticity $(\epsilon_1,\epsilon_2)$ we expect to measure $(-\epsilon_1,-\epsilon_2)$ in the rotated frame. This is however not the case and we find a residual bias of the same order as the additive shear bias that we see in the rest of our analysis. This suggests that this bias is likely to be due to a $\sim 0.01\%$ asymmetry introduced by the shear measurement algorithms. We note however that the shift in the additive bias $\Delta c$ due to the faint galaxies has a significance of about 2$\sigma$ which is much smaller than the effect on the multiplicative bias, which is more than 10$\sigma$, and hence we focus on the latter in the remainder of the paper.

\aaa{We finally stress that our results might be affected by sample variance given the small sample of faint galaxies available in the observations. Calculating Poisson errors on galaxy number counts, we find a statistical variation of $\pm10\%$ in the number of neighbors with $m\leq29$ in the UDF. Including this variation in the simulations, we find an impact on $\Delta\mu$ of the same order as the reported uncertainty. This tends to show that our results are somewhat robust to sample variance. We however only study the impact of sample variance on the number of objects;  the variance in galaxy profiles could also significantly affect the results of this analysis, but this is more difficult to account for without relying on larger observational data sets.}

\subsection{Dependence on the limiting magnitude of faint galaxies}

\begin{figure}
\centering
\includegraphics[width=0.5\textwidth,clip]{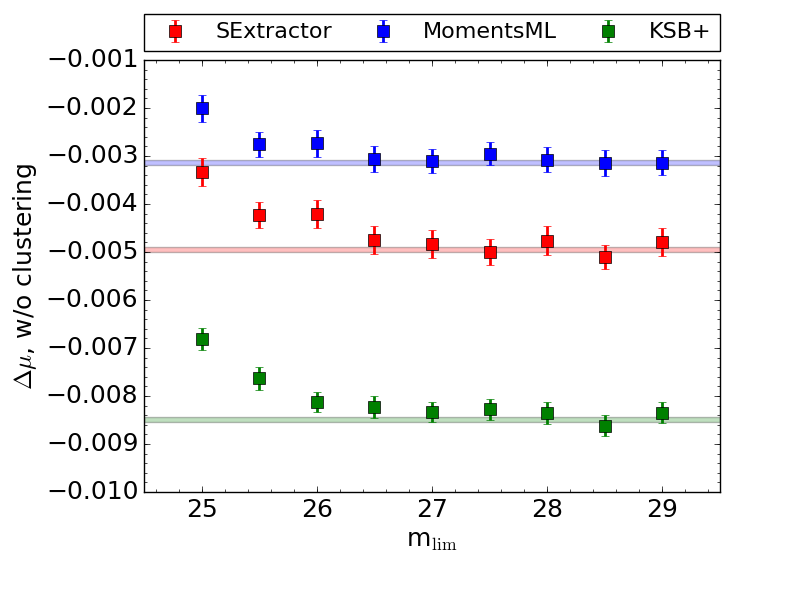}
\caption{Shift in the multiplicative bias due to the presence of the faint galaxies up to the limiting magnitude given on the $x$-axis, without taking the clustering of the faint galaxies into account. Red, blue, and green squares represent the \texttt{SExtractor}/\texttt{PSFEx}, \texttt{MomentsML}, and \texttt{KSB+} measurements, respectively. The shaded regions correspond to a \aaa{goal accuracy of a} $10^{-4}$ variation in $\mu$, centered between the values of the two faintest limiting magnitudes for each method. Every point corresponds to 20 million shear measurements.}
\label{fig:noclus}
\end{figure}

In Fig.~\ref{fig:noclus} we display $\Delta\mu$ for the three methods as a function of the limiting magnitude of the faint galaxies included in the simulations, with shaded areas to show the variation of $\Delta\mu=10^{-4}$. This latter value corresponds to the accuracy that we aim to achieve in the simulations so as not to introduce any bias to the total accuracy requirement of $2\times10^{-3}$. Unsurprisingly, we find that the brighter the faint galaxies, the larger the impact they have on the shear measurement of the bright galaxies. We also see that the shift in the multiplicative bias converges with the magnitude limit of the faint galaxies for each different method. Once again, \texttt{MomentsML} seems to be less affected than the other methods, followed by \texttt{SExtractor} and then \texttt{KSB+}. \texttt{MomentsML} and \texttt{SExtractor} show a change in the multiplicative bias that converges to variations  of less than $10^{-4}$ at magnitude $m=26.5$, and \texttt{KSB+} at $m=27.0$. We conclude that faint galaxies at least brighter than magnitude 26.5 (and even 27 if we want to be inclusive regarding the three tested methods) must be included in the simulation of calibration to avoid biasing shear values of the order of a few times $10^{-3}$, with an uncertainty on this bias of $\sim2\times10^{-4}$.

These results are comparable to those of \citet{Hoekstra+17}. These latter authors found that the faint galaxies induce a negative multiplicative bias of a few times $10^{-3}$ for their KSB shear measurement algorithm. This corresponds to a bias with the same sign and order of magnitude  as that in our study, but one that is a factor of two smaller than that of our KSB implementation. Their Figure 7 also shows a bias that continues to increase up to $m\sim28.5$, which is fainter than the result we obtain with our KSB method. These results also depend on the background determination as shown in Figure 11 of \citet{Hoekstra+17}. There are several differences between their study and ours that could explain this dissimilarity. They use a different rms for the background Gaussian noise and also a slightly smaller dispersion of the intrinsic ellipticity ($\sigma_\epsilon=0.25$). The main difference however is in the simulation of the faint galaxy population. \citet{Hoekstra+17} extrapolate the magnitude and size distributions of the faint galaxies from measurements of GEMS \citep{Rix+04} galaxies with $20<m<25$. In contrast, we measure these properties in the deeper UDF images and therefore include a more realistic population of faint galaxies up to $m=29$. Basing the faint galaxy properties on the bright ones overestimates the number of faint galaxies above magnitude 26 and underestimates the size of these galaxies (see Figures 1 and 2 of \citet{Hoekstra+17} and Fig.~\ref{fig:rmag} of the present analysis). This means that in \citet{Hoekstra+17} there are more faint galaxies than in the present study and that they spread over fewer pixels. Although the galaxy density increase should increase the effect of the faint galaxies, the effect of the change in size is more difficult to predict. These latter authors tested for this effect by changing the size of the faint galaxies by a multiplicative factor and found that a decrease in galaxy size results in an increased impact of the faint galaxies. The two effects mentioned therefore tend to increase the impact of the faint galaxies in \citet{Hoekstra+17}, which could be why they found a higher sensitivity to galaxies fainter than magnitude $27$ than we do.

In addition, we find that the impact of the faint galaxies measured with KSB is highly dependent on the radius of the Gaussian weight function, which is employed to compute galaxy brightness moments \citep[see also Table 1 of][]{Hoekstra+17}. \aaa{Quantitatively, increasing the radius of the weighting function by a factor 1.5 produces a multiplicative bias shift of $\Delta\mu^{\rm KSB}=(-18.34\pm0.40) \times 10^{-3}$ when including faint galaxies up to $m=29$, which corresponds to a change of $\sim 10^{-2}$ compared to our fiducial case. We note also that the error bars increase as the shape estimates are noisier when the radius of the weighting function gets larger compared to the separation between bright and faint galaxies.} This means that different KSB methods are likely to have different sensitivity to the faint galaxies. Adapting the radial weighting function could be a promising way of mitigating the impact of the faint galaxies with KSB measurements, although testing such mitigation procedures is beyond the scope of this paper. Since the weight function is usually chosen to maximize the S/N, such an approach would also introduce a change in the noise bias.

\subsection{Importance of using measured properties for the faint galaxies}
\label{sec:rad}

We further investigate the impact of the size of the faint galaxies on the multiplicative bias shift. In particular we want to know whether one can extrapolate faint galaxy sizes from bright ones, or if one should rather use observed sizes for the faint galaxies, as we do in the rest of the paper.

To test this, we \corr{run} another set of simulations where the sizes of the faint galaxies \corr{are} computed from the extrapolation of the bright galaxy sizes as shown in Fig.~\ref{fig:rmag}. Faint galaxies now \corr{appear} smaller by a factor that is the ratio of the mean half-light radius in each magnitude bin to the extrapolated mean half-light radius for the same magnitude bin. For $m>27$, the extrapolation of the sizes becomes negative and we therefore \corr{do} not include fainter galaxies in this test.

The new multiplicative bias shifts, from 20 million shear measurements, are shown in Table~\ref{tab:1} for \texttt{SExtractor}, \texttt{MomentsML}, and \texttt{KSB+}. We see that decreasing the size of the faint galaxies increases their impact for all methods, confirming the trend observed in \citet{Hoekstra+17}. In our study the shift is lower by about \corr{$0.5$ to $1\times 10^{-3}$} compared to the value measured with the real faint galaxy sizes up to $m=29$. We note that this shift is also significant compared to the fiducial case up to $m=27$, showing that the difference is dominated by the size reduction of the faint galaxies due to the extrapolation and is only marginally affected by the missing galaxies fainter than $m=27$.

These results demonstrate that galaxy sizes should be based on measured galaxy properties and cannot be extrapolated from galaxies brighter than the VIS magnitude limit of 24.5. However, this does not necessarily mean that galaxy sizes need to be measured up to $m=29$. We however do not try to find the minimum depth that would allow one to perform an accurate extrapolation because of the small sample of observed galaxies in this study.

\begin{table}
   \caption{Shifts in the shear multiplicative bias due to the faint galaxies. The column headed ``fiducial'' corresponds to our standard analysis (Sect.~\ref{subsec:bias}), ``radius'' to the case where faint galaxy sizes are extrapolated from that of bright ones (Sect.~\ref{sec:rad}), and ``background'' to the case without background subtraction (Sect.~\ref{sec:bk}). Faint galaxies are included up to magnitude $m=29$, except for the ``radius'' case where faint galaxies are included only up to magnitude $m=27$, since the size extrapolation reaches zero for fainter objects.}
  \centering
 \begin{tabular}{cccc}
   \hline
   \hline 
   &fiducial& radius & background \\
  \hline 
 w/o clustering  & &  &  \\

  $\Delta\mu^{\rm SEx} \times 10^{3}$   &  $ -4.79 \pm 0.30 $  &  $ -5.71 \pm 0.30 $  &  $ -6.78 \pm 0.26 $ \\
  $\Delta\mu^{\rm ML} \times 10^{3}$     &  $ -3.14 \pm 0.27 $  &  $ -3.40 \pm 0.26 $  &  $ -3.69 \pm 0.24 $ \\
  $\Delta\mu^{\rm KSB} \times 10^{3}$   &  $ -8.35 \pm 0.21 $  &  $ -9.03 \pm 0.23 $   &  $ -9.17 \pm 0.23 $ \\
   \hline 
   \hline
 \end{tabular}
 \label{tab:1}
 \end{table}

\subsection{Effect of the background subtraction}
\label{sec:bk}

In this section we decipher the impact of the background subtraction on the shift in the multiplicative bias due to the faint galaxies. In the rest of the paper, the mean background is computed with {\texttt SExtrator} for every patch of $64\times64$ pixels and subtracted from the image before measuring the shear. Here, we do not subtract the background and assume that the mean background is equal to zero for every method. Although the source-free background is indeed equal to zero by construction in our simulations, the effective mean background is slightly higher in the simulations that include the faint galaxies because they contribute a positive noise on top of the sky background.

This measurement leads to the multiplicative bias shifts between the cases with bright and faint galaxies and that with bright galaxies only, shown in the fourth column of Table~\ref{tab:1}. The faint galaxies have been included up to magnitude 29 for 20 million bright galaxies and this can be compared to the same simulations where measurements are taken after subtracting the background estimated by \texttt{SExtractor} in Sect.~\ref{subsec:bias} (second column of Table~\ref{tab:1}). Not subtracting the background increases the absolute value of the shift by about 10 to 30\% for all three measurement methods. We note that in \citet{Hoekstra+17} this shift is of the same order, but with the opposite sign. The comparison between both studies is however difficult in that case, since we position bright galaxies isolated on a grid, while \citet{Hoekstra+17} positioned them randomly with possible blends, resulting in very different background estimates. Both studies agree that it is important to account for the faint galaxies in the background when measuring the shear. But even more important, this shows that any shear measurement strongly depends on the treatment of the background, which can induce multiplicative biases of the order of a few times $10^{-3}$ when including faint galaxies.

\section{Impact of faint galaxy clustering}
\label{sec:clus}

In contrast to the previous section, we now position the faint galaxies according to their clustering around the bright ones. The clustering properties are measured on the UDF images and are described in Sect.~\ref{sec:method}. The simulations remain the same as before, changing only the positions of the faint galaxies.

\subsection{Dependence on the limiting magnitude of the faint galaxies}

\begin{figure}
\centering
\includegraphics[width=0.5\textwidth,clip]{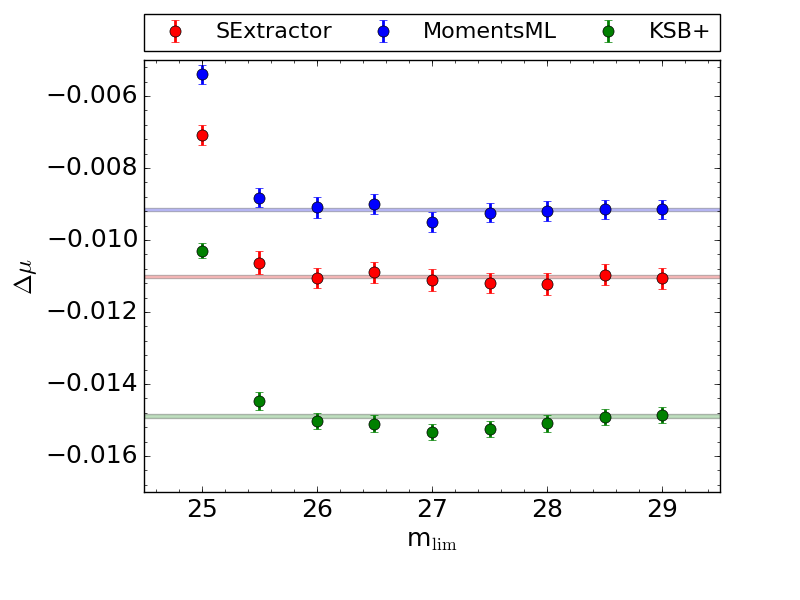}
\caption{Shift in the multiplicative bias due to the presence of the faint galaxies up to the limiting magnitude given on the $x$-axis, when including the clustering of the faint galaxies. Red, blue, and green dots represent the \texttt{SExtractor}/\texttt{PSFEx}, \texttt{MomentsML}, and \texttt{KSB+} measurements, respectively. The shaded regions correspond to a $10^{-4}$ variation in $\mu$, centered between the values of the two deepest limiting magnitudes for each method. Every point corresponds to 20 million shear measurements.}
\label{fig:clus}
\end{figure}

Figure~\ref{fig:clus} shows the shift in the multiplicative bias due to the faint galaxies when they are clustered. We also show shaded regions corresponding to a $10^{-4}$ variation in $\mu$. We see that the impact of the faint galaxies dramatically increases due to the clustering. The shift $\Delta\mu$ is of the order of $10^{-2}$ which is about two to three times larger than when the clustering is not included and two orders of magnitude larger than the accuracy required in the {\it Euclid} calibration simulations. The clustering places faint galaxies closer to the bright ones, which intensifies their impact. Clustering therefore needs to be included in the simulations for the calibration of shape measurement algorithms.

In this case, most of the effect is driven by galaxies of magnitude brighter than 26.5, although the change at fainter magnitude remains significant (i.e., greater than $10^{-4}$) up to magnitude 27.5 for \texttt{MomentsML} and 28 for \texttt{KSB+}. The value of the bias also differs between methods. At magnitude 29, the least affected method is still \texttt{MomentsML}, with a shift of $\Delta\mu^{\rm ML}=(-9.15\pm0.27) \times 10^{-3}$, followed by \texttt{SExtractor} with $\Delta\mu^{\rm SEx}=(-11.06\pm0.29) \times 10^{-3}$ and \texttt{KSB+} with $\Delta\mu^{\rm KSB}=(-14.87\pm0.22) \times 10^{-3}$. According to these results, faint galaxies need to be included in the calibration simulations at least up to magnitude 26.5, and up to 28 for the most affected methods, \corr{and including proper clustering properties}.

\aaa{Propagating Poisson errors on the number of faint neighbors measured in the UDF produces a variation of up to three times the uncertainty on $\Delta\mu$. This is a larger impact than in the case without clustering, for the same reason that the angular separation between bright and faint galaxies is smaller on average when including the clustering, increasing the overall impact of the faint galaxies.}

\subsection{Dependence on clustering length}

In the previous subsection we showed that the clustering of faint galaxies around bright ones has a major impact on shear measurements. Since this is such an important effect, we now try to characterize how well we need to know the clustering, and in particular to what separation from the bright galaxies, referred to as the ``clustering length'' $\theta_{\rm lim}$, it should be accounted for.

In contrast to the rest of the paper, where $\theta_{\rm lim}$ is set to $3\arcsec$, we now vary it from $1\arcsec$ to $3\arcsec$ in steps of $0''\!\!.5$. This means that we include faint galaxies only within the given clustering length, and reject all faint galaxies that would be further away from their bright neighbors. For this test, the magnitude limit of the faint galaxies is set to 29 to make sure we include the effect of all faint galaxies, although we showed in the last section that the multiplicative bias converges for slightly brighter magnitude limits.

\begin{figure}
\centering
\includegraphics[width=0.5\textwidth,clip]{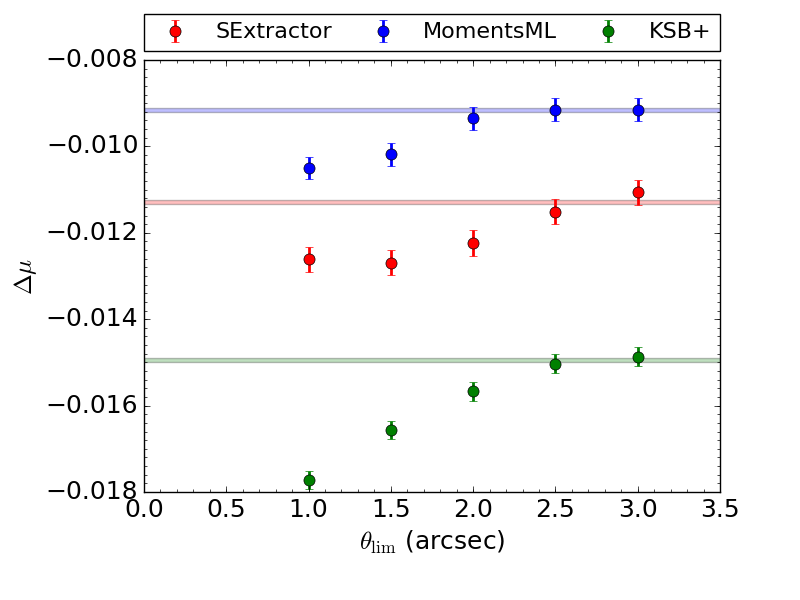}
\caption{Shift in the multiplicative bias due to the presence of the faint galaxies with magnitude brighter than 29 and up to the limiting clustering length $\theta_{\rm lim}$ given on the $x$-axis. Red, blue, and green dots represent the \texttt{SExtractor}/\texttt{PSFEx}, \texttt{MomentsML}, and \texttt{KSB+} measurements, respectively. The shaded regions correspond to a $10^{-4}$ variation in $\mu$, centered between the values of the two largest clustering lengths for each method.}
\label{fig:cl1}
\end{figure}

The results are shown in Fig.~\ref{fig:cl1} for the three methods, together with shaded regions corresponding to a $10^{-4}$ variation in the multiplicative bias. In this figure, each point has been computed from 20 million shear measurements. We find that the bias is larger in absolute value when the clustering length is smaller, for all methods. This result appears surprising at first sight, since it means that the fewer galaxies we include, the more biased we are. However, this can be understood in terms of mitigation of the large impact of close galaxies by those further away from the bright one. Since faint galaxy position angles are uncorrelated, additional faint galaxies are more likely to stretch the bright galaxy or to affect the local background in a direction that is different from that of the closer faint galaxies, partially compensating for their effect. All three methods present a multiplicative bias shift that seems to no longer vary above $\theta_{\rm max}=2''\!\!.5$. This statement is based only on the last two points at $\theta_{\rm max}=2''\!\!.5$ and $\theta_{\rm max}=3''$ and some points at larger radii would be necessary to secure the convergence of the multiplicative bias shift with clustering length. This is however in qualitative agreement with Fig.~\ref{fig:radpermag}, which shows that the excess density of galaxies is significant up to the same clustering length. This tends to show that it is important to include faint galaxies at least up to $\theta_{\rm lim}=2''\!\!.5$ and probably even $3\arcsec$, although more tests will be needed in order to define this value robustly.

\begin{figure}
\centering
\includegraphics[width=0.5\textwidth,clip]{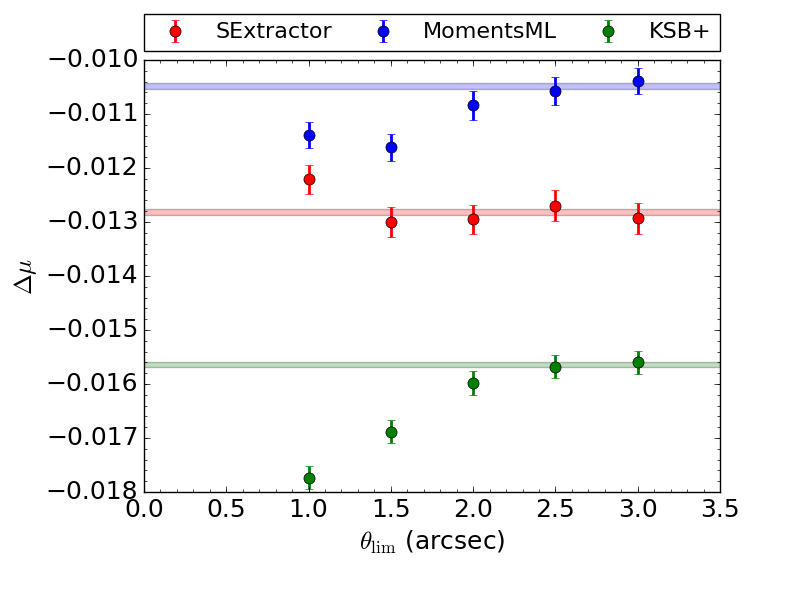}
\caption{Same as Fig.~\ref{fig:cl1}, but with the mean background set to zero in all three measurement methods.}
\label{fig:cl2}
\end{figure}

The effect that we see in Fig.~\ref{fig:cl1} could also be attributed to the background estimate. We test this by remeasuring the shear in these simulations without subtracting the background, as in Sect.~\ref{sec:bk}, and display these results in Fig.~\ref{fig:cl2}. For \texttt{MomentsML} and \texttt{KSB+} we see a similar effect as with the background subtraction, but with slightly larger absolute biases, since faint galaxies are no longer accounted for in the background estimate. For \texttt{SExtractor} however, the multiplicative bias converges for a clustering length of $1''\!\!.5$ and seems almost constant across the full range of $\theta_{\rm lim}$ when we do not subtract the background. This suggests that model-fitting methods are better at dealing with blends and are therefore less affected by what happens further away from the studied object,  meaning that the changes in the bias as a function of the clustering length are driven by the effect on the background estimate for \texttt{SExtractor}. There is no such obvious conclusion for the two other methods. Background subtraction is a more realistic approach for shear measurement, and therefore a clustering length of at least $2''\!\!.5$ should be retained.

\subsection{Impact of the deblending strategy}
\label{sec:deb}

In this section we ascertain the impact of the deblending strategy used to measure the clustering of the faint galaxies in the UDF images. For the main results of the paper we use deblending sub-thresholds {\small DEBLEND\_NTHRESH} and a minimum-contrast parameter for deblending {\small DEBLEND\_MINCONT} of 16 and 0.01, respectively. Here we test two additional deblending schemes, an aggressive deblending with a {\small DEBLEND\_NTHRESH} of 32 and a {\small DEBLEND\_MINCONT} of 0.001, referred as the ``strong-deblending'' case, and a less-aggressive one with a {\small DEBLEND\_NTHRESH} of 8 and a {\small DEBLEND\_MINCONT} of 0.05, referred as the ``weak-deblending'' case. These two additional setups allow us to probe the full range of deblending parameters recommended in the \texttt{SExtractor} documentation: {\small DEBLEND\_NTHRESH} between 8 and 32, and {\small DEBLEND\_MINCONT} between 0.001 and 0.01. Our weak-deblending case is even outside the recommended {\small DEBLEND\_MINCONT} range, to verify whether a minimal deblending strategy still leads to a bias.

\begin{figure}
\centering
\includegraphics[width=0.45\textwidth,clip]{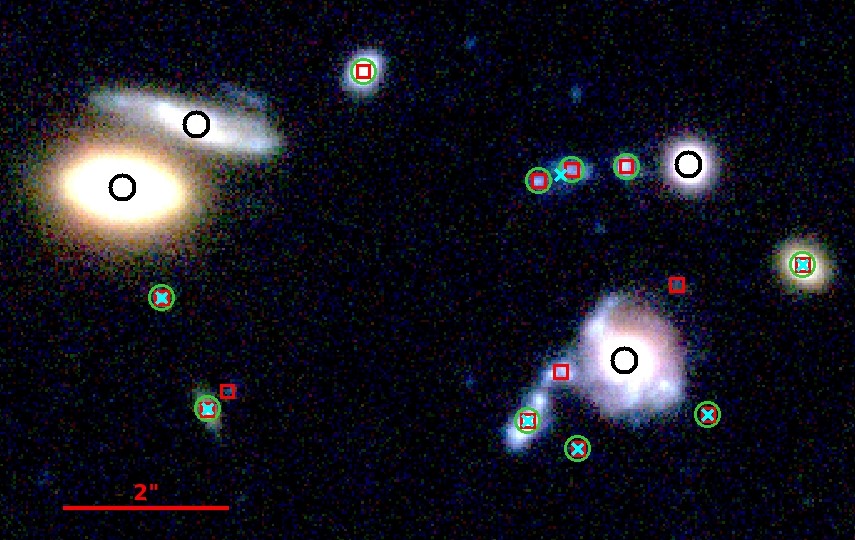}
\includegraphics[width=0.45\textwidth,clip]{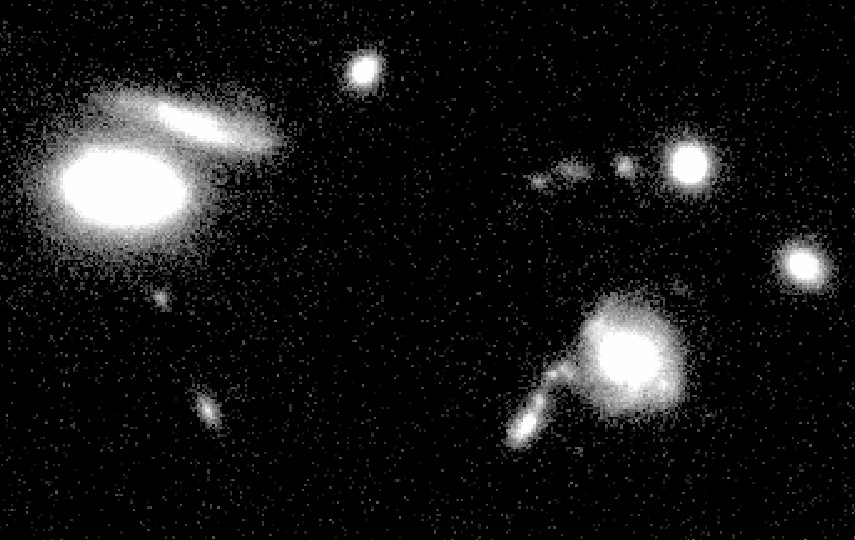}
\includegraphics[width=0.45\textwidth,clip]{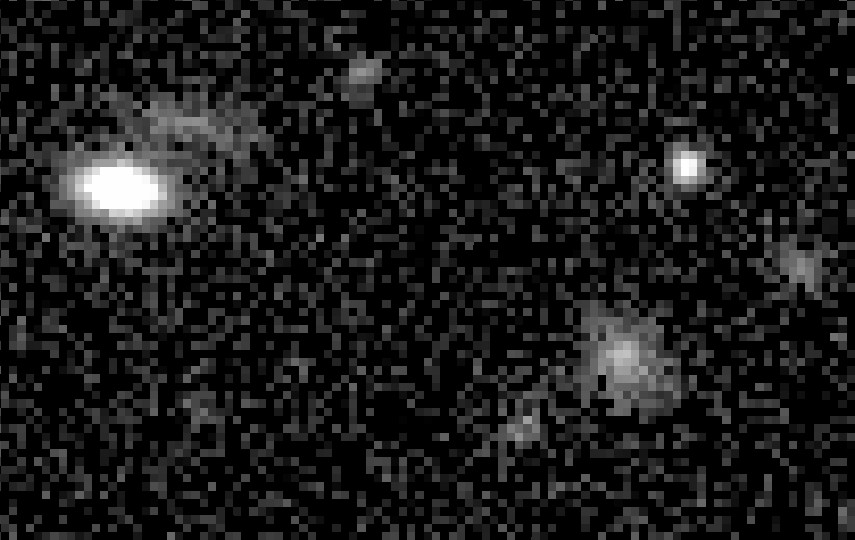}
\caption{HST UDF color image with galaxies brighter than $F775W=24.5$ circled in black, and faint galaxies within $3\arcsec$ of a bright one marked by a cyan cross, green circle, or red square, for weak-, fiducial-, and strong-deblending schemes, respectively ({\it top}). The {\it middle panel} shows the F814W UDF image only and the {\it bottom panel} the expected VIS image computed by degrading the GOODS-South image in the same filter.}
\label{fig:db}
\end{figure}

We recall that the strong-deblending strategy will detect most faint satellite galaxies at the cost of also detecting star-forming regions as faint galaxies, while the weak-deblending case will miss some of the faint satellite galaxies. This is illustrated in the UDF color image shown in the top part of Fig.~\ref{fig:db}, where faint neighbors are marked with a cyan cross when they are detected with weak deblending parameters, with a green circle when detected with fiducial deblending, and a red square with strong deblending. Black circles represent bright galaxies in the fiducial deblending case. We see in particular that the strong-deblending strategy allows us to recover some faint blends, but detects several star-forming regions in the spiral galaxy in the lower right corner of the image. The weak-deblending strategy misses several faint galaxies, such as the one in the top of the image, but is less affected by star-forming regions. The fiducial case is a mixture of the two others, being affected by a few spiral substructures while recovering most of the faint satellites.

To better distinguish between the impact of star-forming regions and faint galaxies, we should frame the discussion in terms of what {\it Euclid} will see. In Fig.~\ref{fig:db}, we therefore also show the F814W UDF image of the same region of the sky in the middle panel and the equivalent VIS image in the bottom panel. The latter has been computed by applying a re-binning from $0''\!\!.03$ to $0''\!\!.1$ and a re-convolution from the HST PSF to the expected VIS PSF. This is based on the F814W images of the GOODS-South survey \citep{Giavalisco+04}. Since GOODS is shallower than UDF, most galaxies fainter than magnitude about 27 are not included in the original image, but this is unimportant for the present discussion, since these galaxies disappear given the noise level of the {\it Euclid} VIS-like image. We see that without color, some star-forming regions are already difficult to identify in the F814W UDF image, and that identification of star-forming regions becomes impossible in the VIS emulated image. In particular the potential star-forming region or merger in the bright galaxy of the lower right corner now appears as a separate very faint object. This tends to validate our approach of treating star-forming regions and faint clustered galaxies in the same way in this study, since they will not be disentangled from another in the {\it Euclid} VIS images.

We now build two new additional sets of simulations, corresponding to the two other deblending strategies. All galaxy parameters are affected by the deblending strategy: not only the number of faint neighbors and their positioning around the bright galaxies, but also the fluxes, half-light radii, and S\'ersic indices of the bright and faint galaxies. The more aggressive the deblending, the stronger the effect expected on the shear, as it will mean more faint galaxies closer to the bright ones. We include faint galaxies up to magnitude 29 and use a clustering length of $3\arcsec$. Given the high computational cost, we do not study the convergence of the bias with the limiting magnitude of the faint galaxy sample in this case. Results are displayed in Table~\ref{tab:db} for all three methods and for the two cases where galaxies are randomly positioned and where they follow the clustering measured in the UDF with the different deblending strategies.

\begin{table}
   \caption{Shifts in the shear multiplicative bias due to the faint galaxies with density and clustering measured on the UDF data for various deblending strategies. Weak deblending refers to ({\small DEBLEND\_NTHRESH}, {\small DEBLEND\_MINCONT}) values of (8, 0.05), fiducial deblending to (16, 0.01), and strong deblending to (32, 0.001)}
  \centering
 \begin{tabular}{cccc}
   \hline
   \hline 
   &weak& fiducial & strong\\
   &deblending & deblending &deblending \\
  \hline 
 w/o clustering  & &  &  \\

  $\Delta\mu^{\rm SEx} \times 10^{3}$   &  $ -4.91 \pm 0.28 $  &  $ \,\,\,-4.79 \pm 0.30 $  &  $ \,\,\,-8.27 \pm 0.28 $ \\
  $\Delta\mu^{\rm ML} \times 10^{3}$     &  $ -2.63 \pm 0.27 $  &  $ \,\,\,-3.14 \pm 0.27 $  &  $ \,\,\,-6.50 \pm 0.28 $ \\
  $\Delta\mu^{\rm KSB} \times 10^{3}$   &  $ -8.20 \pm 0.22 $  &  $ \,\,\,-8.35 \pm 0.21 $  &  $ -11.30 \pm 0.23 $ \\
 with clustering  & &  &  \\
  $\Delta\mu^{\rm SEx} \times 10^{3}$   &  $ -3.99 \pm 0.31 $  &  $ -11.06 \pm 0.29 $  &  $ -36.98 \pm 0.35 $ \\
  $\Delta\mu^{\rm ML} \times 10^{3}$    &  $ -2.20 \pm 0.29 $  &  $ \,\,\,-9.15 \pm 0.27 $  &  $ -35.29 \pm 0.30 $ \\
  $\Delta\mu^{\rm KSB} \times 10^{3}$   &  $ -7.16 \pm 0.21 $  &  $ -14.87 \pm 0.22 $  &  $ -43.26 \pm 0.26 $ \\
   \hline 
   \hline
 \end{tabular}
 \label{tab:db}
 \end{table} 

In the weak-deblending case the multiplicative bias shift is consistent with that of the fiducial deblending, for all three methods, when galaxies are randomly positioned. This means that the change in the density of faint neighbors in this case is small enough compared with the fiducial approach, although the variation in the separation might still be significant. When including clustering, the change in the shift compared to the random positioning is less dramatic than in the fiducial case, with a change of about $+0.5$ to $+1.0\times 10^{-3}$ compared to about $-6\times 10^{-3}$. The fact that it is a positive difference means that with the weak-deblending strategy we detect fewer faint clustered galaxies than in the field. This is expected: with very weak deblending, faint clustered galaxies are not separated from the bright ones. Although the impact of clustering is lower in this case, it is an order of magnitude higher than the accuracy we want to achieve in the calibration simulations ($10^{-4}$), meaning that clustering would still need to be accounted for even with such an unrealistically weak deblending strategy. These results might however depend on the complexity of the galaxy modeling, since our single-S\'ersic model approach does not account for galaxy substructures that are included in the shapes of the bright galaxies  with weak deblending.

The strong-deblending case shows the opposite behavior compared to weak-deblending. The shifts in the multiplicative biases due to the faint galaxies are strongly increased compared to the fiducial case. The change in the absolute bias increases by about 25 to 50\% when galaxies are randomly positioned and by more than a factor of three when they are placed according to their clustering. With this strategy, a higher number of faint objects is detected, especially in the close vicinity of bright ones, since these faint structures also correspond to star-forming regions. As in the weak-deblending case, this result can also be interpreted in terms of morphology: the bright galaxy shapes are better modeled by a single-S\'ersic profile in the strong-deblending case and substructures are now included in the faint galaxy population.

Through these two additional sets of simulations, we see that the deblending strategy used to measure the clustering of the faint galaxies has a significant impact on the shift in the multiplicative shear bias due to the inclusion of faint galaxies. This is seen for all ranges of possible deblending parameters; it will therefore be very important to find a consistent way to \corr{implement the deblending in} the calibration simulations of {\it Euclid}. This problem is also linked to the issue of including real galaxies in the calibration simulations, since faint close galaxies and star-forming regions would probably be included in the observed patches that would be passed on into the simulations.

\section{Impact of magnification}
\label{sec:magnif}

In the observations, faint background galaxies are magnified due to the presence of the bright foreground objects along the line of sight. This will have two main effects: the faint galaxies will appear brighter and shifted from their original positions, and some fainter galaxies that were not detected before magnification will become detectable. Both effects should affect the correlated noise, and the goal of this section is to ascertain the impact on the bright galaxy shape measurement. The amplification of intensity and the appearance of fainter galaxies will increase the impact of the faint galaxies, whereas the shift in position will enlarge the separation between bright and faint galaxies, decreasing the impact of the latter on shape measurement. In this first study, we neglect the appearance of magnified galaxies, which is a weaker effect as it concerns only the faintest galaxies. This approximation is justified by the fact that the multiplicative bias asymptotes to a certain value when we further add galaxies above $m\sim28,$ regardless of the measurement method used (Figs.~\ref{fig:noclus} and \ref{fig:clus}). Magnification also changes the properties of the faint galaxies as a function of separation from the bright ones, which is currently ignored. Our goal here is to check whether magnification can safely be neglected or if it is a major effect that needs to be modeled. To this end we implement an approximate artificial magnification of the faint galaxies through position shift and intensity amplification.

We apply magnification in two different ways. The first approach is to consider that all faint galaxies are behind the bright ones. In this case we place the faint galaxies at a source redshift of $z_{\rm s}=2$ and the bright galaxies at a lens redshift of $z_{\rm l}=1$. These numbers roughly correspond to the expected median redshifts of these populations. In a more refined approach, we assign a redshift to every simulated galaxy. We draw a redshift for every bright galaxy, using the photometric redshift probability distribution function measured in our UDF galaxy sample cross-matched with the photometric redshift catalog of \citet{Rafelski+15}. For the faint galaxies, we do not draw from the redshift distribution, but from the distribution of the redshift difference between bright and faint galaxies. This is to preserve the correlation between lens and source redshifts, which is important for lensing. The measured and generated distributions of redshifts are displayed in Figs.~\ref{fig:histoobs} and \ref{fig:histocat}, respectively. We see that the faint redshift distribution is slightly distorted due to our choice of preserving the redshift separation between bright and faint galaxies instead of using the faint galaxy redshift distribution. Although this second approach is expected to be more accurate, the method with fixed source and lens redshift planes does not rely on the photometric redshift measured by \citet{Rafelski+15} and is therefore an interesting cross-check.

In both cases, we calculate the magnification using a spherical Navarro, Frenk, and White mass density profile \citep{NFW97}. Every bright galaxy is taken to have a mass of $M_{200{\rm c}}=5.00 \times 10^{11}\,h^{-1} {\rm M_{\odot}}$ and a concentration parameter of $c_{200{\rm c}}=5.5$, corresponding to the value for halos of this mass at $z=1$ in \citet{Gao+08}. The virial radius is set to $r_{200}=69~{\rm kpc}$, which is computed from the mass and concentration parameters with $h=0.7$. We calculate the corresponding shear modulus and convergence at the positions of the faint galaxies using the analytical profile from \citet{Wright+00}. We apply the flux magnification using the `galsim.magnify' \texttt{GalSim} function and also shift positions accordingly to the induced shear and convergence values, using the Jacobian matrix of the lensing transformation. The mean magnification applied to faint background galaxies is $\sim1.2$ and $\sim1.05$, when using fixed redshift planes and the photometric redshift distribution, respectively.

\begin{table}
   \caption{Shifts in the shear multiplicative bias due to the magnification of faint galaxies (up to magnitude 29) by bright foreground ones. ``magnif. 1'' corresponds to the case where the magnification is computed while considering bright galaxies to lie in the redshift plane $z_{\rm l}=1$ and faint galaxies in the redshift plane $z_{\rm s}=2$. ``magnif. 2'' corresponds to the case where the magnification is computed using redshift distributions sampled from the photometric-redshift catalog of \citet{Rafelski+15}.}
  \centering
 \begin{tabular}{cccc}
   \hline
   \hline 
   &fiducial           & magnif. 1   & magnif. 2\\
   &(w/o magnif.) & (z planes) & (z distribution)\\
  \hline 
 w/o clustering  & &  &  \\

  $\Delta\mu^{\rm SEx} \times 10^{3}$     &  $ -4.79 \pm 0.30 $  &  $ -4.93 \pm 0.28 $ &  $ -4.93 \pm 0.30 $ \\
  $\Delta\mu^{\rm ML} \times 10^{3}$      &  $ -3.14 \pm 0.27 $  &  $ -3.03 \pm 0.27 $ &  $ -3.02 \pm 0.27 $ \\
  $\Delta\mu^{\rm KSB} \times 10^{3}$     &  $ -8.35 \pm 0.21 $  &  $ -8.51 \pm 0.22 $ &  $ -8.36 \pm 0.23 $ \\
   \hline 
   \hline
 \end{tabular}
 \label{tab:magnif}
 \end{table} 
 
The effect of magnification is shown in Table~\ref{tab:magnif} for 20 million shear measurements and including faint galaxies up to magnitude $m=29$. We investigate only the case without including the clustering of the faint galaxies. This is because only true background galaxies are affected by magnification, and not the clustered faint galaxies at the redshift of the bright galaxy. The resulting shift in the multiplicative bias is about $10^{-4}$ for both magnification calculations and for any method, which is below the precision of $2$ to $3\times10^{-4}$ that we reach on $\Delta\mu$. 

We conclude that magnification is a secondary effect that can be neglected \aaa{compared to the multiplicative bias total error budget of $2\times10^{-3}$ allowed for \textit{Euclid}}. We note however that our simple magnification model could be improved by using a mass--concentration relation and the galaxy measured radius to assign individual NFW profiles to each galaxy, and by including the correct dependence of magnification on the faint-to-bright galaxy separation. However, we do not try to refine our model, since magnification effects are almost negligible in the present approach.

\section{\corr{Towards the inclusion} of faint galaxies in \textbf{\textit{Euclid}} calibration simulations}
\label{sec:discu}

\aaa{We showed that faint galaxy clustering can cause a percentage-level bias if unaccounted for in calibration simulations. It is therefore mandatory to include it for methods relying on such calibrations.} 
This can be done in two different ways, relying either on observations, as in the present study, or on simulations. We stress however that both approaches should be explored, since they lead to very specific, different systematic biases while accounting for the faint galaxy clustering issue.

\begin{figure}
\centering
\includegraphics[width=0.45\textwidth,clip]{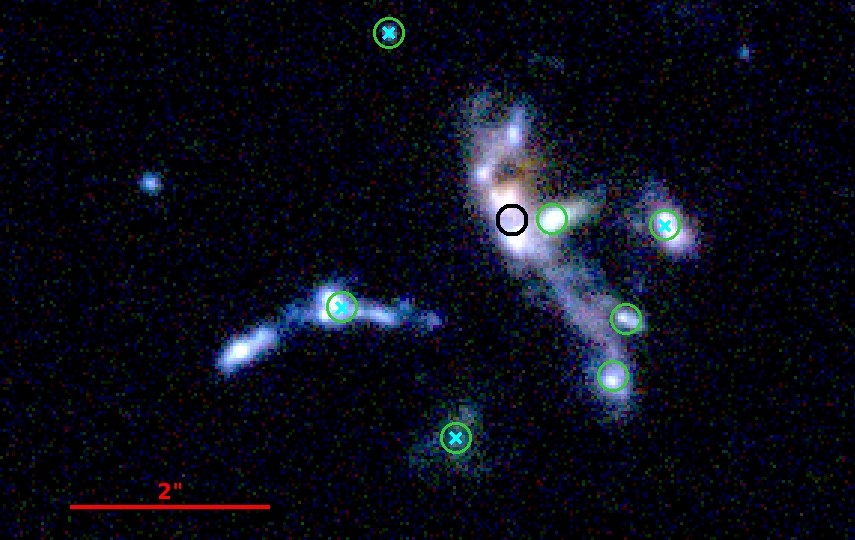}
\includegraphics[width=0.45\textwidth,clip]{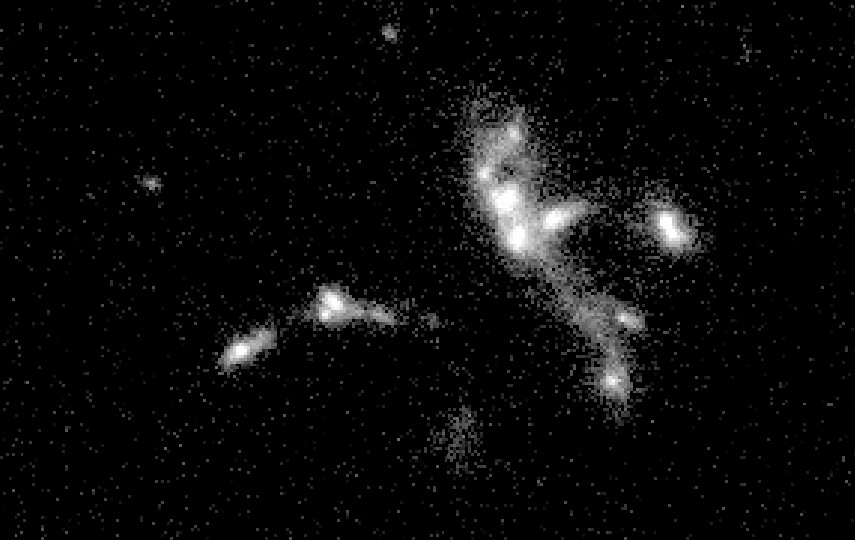}
\includegraphics[width=0.45\textwidth,clip]{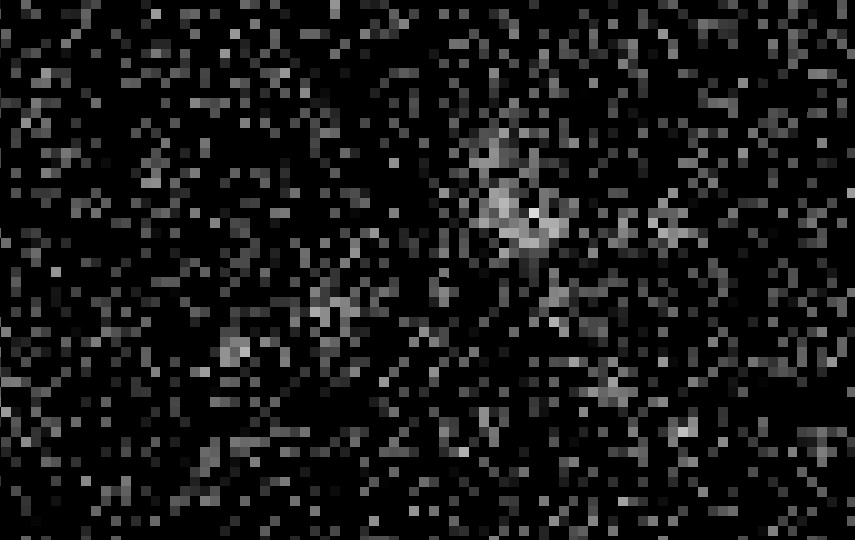}
\caption{Same as Fig.~\ref{fig:db}, centered on an irregular galaxy.}
\label{fig:db2}
\end{figure}

The first method would be to measure the clustering of faint galaxies in a large collection of deep HST data and to include it in the calibration simulations for all magnitudes up to 28. That is the approach we followed in this paper to quantify the impact of the faint galaxy clustering, but using only the UDF data as the observational sample. Although this is the most straightforward method for including faint galaxy properties, we find that the actual value of the measured bias strongly depends on the deblending strategy that is used to measure the number and separation of faint neighbors around bright galaxies. However, by studying a broad range of deblending strategies, we find that the effect of the faint galaxy clustering is significant in all cases. More thought is required on how best to include deblending in the design of the calibration simulations. This will also depend on the deblending method that is used for the detection in the {\it Euclid} observations. This question is also linked to the use of real galaxies instead of simple galaxy models in the simulations. In Fig.~\ref{fig:db2}, we show the UDF color image of an irregular galaxy, and its image in both the F814W filter and the VIS filter. We note the absence of red squares in this image because in the strong deblending case the core of this galaxy is separated into faint objects preventing its classification as a bright galaxy and because we only display faint galaxies with a separation $\theta\leq3\arcsec$ to a bright galaxy. With our fiducial deblending, many substructures of this galaxy are considered as faint clustered objects, which is acceptable since they appear as noise in the equivalent VIS image. If we decide to use observed sky patches instead, these substructures will be directly included in the simulations and this could be a shortcut to the problem of measuring clustering on the smallest scales. This would nonetheless raise some new issues. In particular, the application of the shear in the simulations would be complicated, since it is not possible to have lensing-free training images from HST observations. Likewise, one has to distinguish between the situation of galaxies located at the same redshift and carrying the same shear versus chance projections of galaxies at different redshifts carrying different shear. Such a distinction could however be carried out with deep spectroscopic data, for example the MUSE Ultra Deep Field survey which provides accurate redshifts for galaxies down to magnitude 30 in a subarea of the UDF \citep{Bacon+17,Brinchmann+17}. One also has to account for correlated noise introduced by the shearing, for example by whitening the noise in the extracted images \citep{Rowe+15}.

The second approach would be to position galaxies according to cosmological simulations, which include the clustering properties by construction. These simulations, such as the {\it Euclid} Flagship simulation \citep{Potter+17}, should reach magnitudes down to 26.5, which is close to the limit above which clustered galaxies no longer contribute to the multiplicative bias. Indeed, clustered galaxies are mainly the brightest of the faint sample and typically satellites of the simulated target galaxies. Fainter galaxies could then be placed randomly. To check this possibility we run an extra set of simulations, where galaxies brighter than magnitude 26 follow the clustering properties of the UDF and fainter galaxies are placed randomly up to magnitude 29. Although the Flagship simulation reaches magnitude 26.5, its redshift limit of $z=2.3$ means that about 20 to 25\% of the faint sources are missing. That is why we take a safe estimate of 26 as the magnitude up to which we can obtain a complete sample of faint clustered galaxies from this simulation. Simulating 20 million bright galaxies allows us to derive the change in the multiplicative bias due to the faint galaxies with this particular positioning. We find $\Delta\mu^{\rm SEx}=(-11.56\pm0.28) \times 10^{-3}$, $\Delta\mu^{\rm ML}=(-9.29\pm0.27) \times 10^{-3}$, and $\Delta\mu^{\rm KSB}=(-15.25\pm0.23) \times 10^{-3}$ for the three different measurement methods. These values are close to those obtained when all galaxies are positioned according to their clustering properties. The difference ranges from 1 to 5$\times10^{-4}$ which is close to the acceptable limit for the bias induced by the calibration simulations. This method seems to be a promising way of dealing with the faint galaxy clustering, if we can push the redshift range of the simulation a bit further, allowing us to include slightly fainter galaxies with their clustering properties. This approach however suffers from several other issues. It would require one to first check the accuracy of the clustering in such simulations and to verify that it does not introduce a dependence on the cosmology of the simulation, that is, that the clustering dependence on cosmology has a negligible impact on the shear bias. Another issue is that it will require some understanding of how halos are linked together and in particular how to include realistic star-forming regions and morphologies if we do not rely on observational patches. More generally, the implementation of baryon physics on such small scales is nontrivial and will add further to the uncertainties on the faint galaxy clustering measured from the simulations.

\section{Summary and conclusions}
\label{sec:ccl}

In this paper, we \corr{studied} the impact of the undetected galaxies on the shape measurement of detected galaxies in shear calibration images that resemble those of the {\it Euclid} VIS instrument. We used a realistic sample of galaxies with properties measured in the HST UDF images down to magnitude $m=29$. We further investigated the effect of galaxy clustering and magnification. Shear measurements were performed with three different algorithms, which are representative of methods usually applied in the community.

We confirm the result of \citet{Hoekstra+17} that galaxies need to be included to at least $m=28$ in the calibration images to avoid biasing shear measurement at the order of a few times $10^{-3}$, with an accuracy of $\sim2\times10^{-4}$ on this bias. \aaa{We also find that the shearing of the faint galaxies is an important parameter since it correlates the direction of the
effect of faint galaxies with the input shear applied to the bright galaxies.}

In our simulations there is a significant difference in the multiplicative bias shift due to the faint galaxies ($\sim10^{-3}$) from using faint galaxy sizes extrapolated from the bright sample ($m \leq 24.5$) rather than sizes measured in the UDF. This establishes a need for deep observations to measure the properties of the fainter objects that need to be included in the simulations.

We also show that the clustering of the faint galaxies has a dramatic impact on the multiplicative bias, increasing its value up to $10^{-2}$, and that it must be accounted for at least on scales smaller than $2''\!\!.5$.

However, magnification effects seem to be negligible with our simple implementation, with a change of the order of less than $10^{-4}$ in the multiplicative bias shift. 

Finally, the three measurement methods perform differently and have different levels of sensitivity to the faint galaxies. However, the biases are of the same order and the limiting magnitude to which galaxies need to be included is similar whatever measurement method is used. This is consistent with the fact that the faint galaxy issue is an astrophysical effect and is therefore not associated with a particular shear measurement algorithm. We note that the algorithms used in this work are not designed or tuned to reduce the impact of faint undetected galaxies, while the final {\it Euclid} pipeline will be. As such, the sensitivity we report could be pessimistic, but we do not expect an improvement of the methods to strongly reduce the amplitude of the observed effect.

It is therefore paramount to include the clustering of the faint galaxies in the calibration simulations for the {\it Euclid} mission, and probably also for {\small LSST} and {\small WFIRST}. We have proposed two different ways to achieve this: one based on deep observations, and the other one on cosmological simulations. Each method presents its own strengths and weaknesses and should be further investigated. In particular, the observation-based method is sensitive to the deblending strategy, while the simulation-based method could introduce a cosmological dependence of the shear calibration.

Finally, we stress that the statistics of the faint galaxies are very low in our analysis, given that the UDF dataset represents one of the only available images to reach $m=29$, together with the Frontier field parallels. Although the GOODS survey and later the {\it Euclid} deep field can increase the statistics of galaxies brighter than magnitude 27, further observations will be required in order to measure and include the properties of the faintest galaxies in an \corr{adequate} way. The previous example also shows that it might be sufficient to include the clustering to a less deep magnitude, although it is mandatory to also add randomly positioned fainter galaxies up to magnitude 28. With this approach, deep observations aimed at direct clustering measurements could become less demanding. Extrapolating Poisson errors on galaxy counts from our UDF measurements shows that the {\it Euclid} deep field would be sufficient to characterize the amplitude of the clustering of the faint galaxies up to magnitude 26.5.

\begin{acknowledgements}
We are grateful to Oliver Cordes and Ole Marggraf for their support with computer servers. We thank Raphael Gavazzi and Emmanuel Bertin for useful discussions about the \texttt{SExtractor}/\texttt{PSFEx} shear measurement method, and for making these software packages publicly available. We also thank Patrick Simon, James Bartlett\aaa{, and the anonymous referee} for pertinent suggestions. We are grateful to the \texttt{GalSim} and HST-UDF team for the public release of their software and data, respectively.

NM acknowledges support from the German Federal Ministry for Economic Affairs and Energy (BMWi) provided via DLR under project no. 50QE1103 and from a fellowship of the Centre National d'Etudes Spatiales (CNES). HH acknowledges support from Vici grant 639.043.512, financed by NWO. MT acknowledges support by a fellowship of the Alexander von Humboldt Foundation, and the DFG Emmy Noether grant Hi 1495/2-1. RH acknowledges support from the European Research Council FP7 grant number 279396 and the US Department of Energy under Award Number DE-SC0018053. \aaa{JB acknowledges support by Funda{\c c}{\~a}o para a Ci{\^e}ncia e a Tecnologia (FCT) through national funds (UID/FIS/04434/2013) and Investigador FCT contract IF/01654/2014/CP1215/CT0003., and by FEDER through COMPETE2020 (POCI-01-0145-FEDER-007672).}

The {\it Euclid} Consortium acknowledges the European Space Agency and the support of a number of agencies and institutes that have supported the development of {\it Euclid}. A detailed complete list is available on the {\it Euclid} web site (\texttt{http://www.euclid-ec.org}). 
In particular the Academy of Finland, the Agenzia Spaziale Italiana,
the Belgian Science Policy, the Canadian Euclid Consortium, the Centre
National d'Etudes Spatiales, the Deutsches Zentrum f\"ur Luft- and
Raumfahrt, the Danish Space Research Institute, the Funda\c{c}\~{a}o
para a Ci\^{e}nca e a Tecnologia, the Ministerio de Economia y
Competitividad, the National Aeronautics and Space Administration, the
Netherlandse Onderzoekschool Voor Astronomie, the Norvegian Space
Center, the Romanian Space Agency, the State Secretariat for
Education, Research and Innovation (SERI) at the Swiss Space Office
(SSO), and the United Kingdom Space Agency.
\end{acknowledgements}

\bibliographystyle{aa}
\bibliography{cl}

\end{document}